\documentclass{aastex631}

\usepackage{natbib}
\usepackage{amsmath}
\usepackage{graphicx}
\usepackage{multirow}
\usepackage{hyperref}

\newcommand{\farcsec}{\hbox{$.\!\!^{\prime\prime}$}}

\newcommand{\hr}{HR~4796A}

\newcommand{\iband}{$i'$-band}
\newcommand{\zband}{$z'$-band}
\newcommand{\rband}{$r'$-band}
\newcommand{\gband}{$g'$-band}
\newcommand{\adi}{KLIP-ADI}
\newcommand{\rdi}{KLIP-RDI}

\begin{document}

\title{\texttt{ffortissimo}: A Freeform Forward-Modeling Pipeline for High-Contrast Images of Circumstellar Disks Based on Automatic Differentiation}

\shorttitle{Freeform Forward-Modeling of Disks}
\shortauthors{Kueny et al.}

\author[0000-0001-8531-038X]{Jay K. Kueny}
\affiliation{Steward Observatory, University of Arizona, Tucson, 933 N Cherry Ave, Tucson, AZ 85721, USA}
\affiliation{James C. Wyant College of Optical Sciences, University of Arizona, 1630 E. University Blvd., Tucson, AZ 85721, USA}
\affiliation{National Science Foundation Graduate Research Fellow}

\author[0000-0003-1905-9443]{Joseph D. Long}
\affiliation{Center for Computational Astrophysics, Flatiron Institute, 162 5th Ave, New York, NY}

\author[0000-0002-2346-3441]{Jared R. Males}
\affiliation{Steward Observatory, University of Arizona, Tucson, 933 N Cherry Ave, Tucson, AZ 85721, USA}

\author[0000-0001-6654-7859]{Alycia J. Weinberger}
\affiliation{Earth and Planets Laboratory, Carnegie Institution for Science, 5241 Broad Branch Road NW, Washington, DC 20015-1305}


\author[0000-0002-2167-8246]{Laird M. Close}
\affiliation{Steward Observatory, University of Arizona, Tucson, 933 N Cherry Ave, Tucson, AZ 85721, USA}

\author[0000-0002-4934-3042]{Joshua Liberman}
\affiliation{James C. Wyant College of Optical Sciences, University of Arizona, 1630 E. University Blvd., Tucson, AZ 85721, USA}
\affiliation{Steward Observatory, University of Arizona, Tucson, 933 N Cherry Ave, Tucson, AZ 85721, USA}

\author[0000-0001-5130-9153]{Sebastiaan Haffert}
\affiliation{Leiden Observatory, Leiden University, PO Box 9513, 2300 RA Leiden, The Netherlands}
\affiliation{Steward Observatory, University of Arizona, Tucson, 933 N Cherry Ave, Tucson, AZ 85721, USA}

\author[0000-0003-0843-5140]{Eden McEwen}
\affiliation{Steward Observatory, University of Arizona, Tucson, 933 N Cherry Ave, Tucson, AZ 85721, USA}
\affiliation{James C. Wyant College of Optical Sciences, University of Arizona, 1630 E. University Blvd., Tucson, AZ 85721, USA}
\affil{National Science Foundation Graduate Research Fellow}

\author[0000-0003-3253-2952]{Maggie Y. Kautz}
\affiliation{Steward Observatory, University of Arizona, Tucson, 933 N Cherry Ave, Tucson, AZ 85721, USA}

\author[0000-0002-1097-9908]{Olivier Guyon}
\affiliation{Subaru Telescope, National Observatory of Japan, Hilo, HI}
\affiliation{Steward Observatory, University of Arizona, Tucson, 933 N Cherry Ave, Tucson, AZ 85721, USA}
\affiliation{James C. Wyant College of Optical Sciences, University of Arizona, 1630 E. University Blvd., Tucson, AZ 85721, USA}

\author[0000-0003-3904-7378]{Logan Pearce}
\affiliation{Department of Astronomy, University of Michigan, Ann Arbor, MI}

\author[0009-0005-5534-7495]{Parker T. Johnson}
\affiliation{Steward Observatory, University of Arizona, Tucson, 933 N Cherry Ave, Tucson, AZ 85721, USA}
\affil{National Science Foundation Graduate Research Fellow}

\author[0009-0002-9752-2114]{Katie Twitchell}
\affiliation{Steward Observatory, University of Arizona, Tucson, 933 N Cherry Ave, Tucson, AZ 85721, USA}
\affiliation{James C. Wyant College of Optical Sciences, University of Arizona, 1630 E. University Blvd., Tucson, AZ 85721, USA}

\author[0000-0002-8110-7226]{Jialin Li}
\affiliation{Steward Observatory, University of Arizona, Tucson, 933 N Cherry Ave, Tucson, AZ 85721, USA}
\affil{National Science Foundation Graduate Research Fellow}

\author[0000-0002-5559-1544]{Alex Hedglen}
\affiliation{Northrop Grumman Coroporation, 600 South Hicks Rd, Rolling Meadows, IL}

\author{Avalon Gower}
\affiliation{Draper Laboratory, 555 Technology Square, Cambridge, MA}

\author{Warren Foster}
\affiliation{Steward Observatory, University of Arizona, Tucson, 933 N Cherry Ave, Tucson, AZ 85721, USA}

\author{Jhen Lumbres}
\affiliation{Northrop Grumman in Pasadena, CA}

\author{Lauren Schatz}
\affiliation{Starfire Optical Range, Kirtland Air Force Base, Albuquerque, NM}

\begin{abstract}

Modeling circumstellar disks in the traditional sense carries the assumption that the dust density distribution can be accurately described with a fixed parametric form. Furthermore, commonly-used algorithms for subtracting the stellar point-spread function (PSF) distort the true morphology of the faint underlying disk structure, especially dusty features that are located at small angular separations. These phenomena often lead to significant residuals with parametric disk models and make it difficult to measure the full realizable range of the scattering function of the dust. We address these challenges with \texttt{ffortissimo}, a novel, pixel-based freeform forward modeling pipeline designed to characterize extended objects in KLIP-reduced images. We built this pipeline within the framework of \textsc{Jax}, which is a machine learning library in Python that enables efficient optimization through automatic differentiation (``autodiff") and GPU-accelerated array computations. Using visible light images of the disk around HR~4796A taken by the ``extreme" Magellan Adaptive Optics instrument (MagAO-X), we show that our data-driven freeform models excel at fitting a complex dust distribution and can infer the dust scattering properties even through PSF subtraction artifacts. Additionally, we demonstrate the potential for retrieving spatial dust features beyond the diffraction limit of the telescope. We note that there are remaining challenges to address before precision photometry using these freeform models is advised. These include better background, wind-driven halo, and speckle characterization as preventing the freeform models from learning these noise artifacts is currently difficult.

\end{abstract}


\section{Introduction} 
\label{sec:intro}

Characterizing circumstellar objects through direct imaging is exceptionally challenging as it involves detecting signals at the extreme contrasts. Detecting exoplanets and investigating how they form comprise the main goals in this field and we can achieve the latter through the study of circumstellar disks. Disks are torus or ring-like structures containing the surviving dust and gas in orbit around a star after stellar formation. There are a variety of evolutionary stages for disks with the youngest examples (i.e., protoplanetary disks such as HL Tauri; \citealt{alma_partnership_2014_2015}) acting as planetary nurseries and the oldest (i.e., debris disks such as Fomalhaut; \citealt{gaspar_spatially_2023}) facilitating study of the remnants leftover after planet formation \citep{hughes_debris_2018}. Disks of all evolutionary stages offer unique opportunities to probe what exoplanets are composed of or how they form, which is currently not well understood. In particular, measuring a disk's Scattering Phase Function (SPF) can allow inference of the bulk composition of the scattering material which may offer clues to the composition and evolution of exoplanets (\citealt{arnold_stumbling_2022}; \citealt{shahar_what_2019}). Additionally, complex features in the dust such as gaps, clumps, warps, and spirals can hint toward locations of undiscovered planets or how planets interact with their environments \citep{wyatt_spiral_2005} as was demonstrated with the dusty warp in the Beta Pictoris disk \citep{mouillet_planet_1997}.

The current generation of large ground-based telescopes in the 6- to 10-meter class facilitates unprecedented direct imagery of the circumstellar environment when assisted with adaptive optics (AO). AO instruments such as Magellan/MagAO-X \citep{males_magao-x_2024}, VLT/SPHERE \citep{beuzit_sphere_2019}, Gemini/GPI (\citealt{macintosh_gemini_2008}; \citealt{perrin_polarimetry_2015}), Keck/KPIC \citep{delorme_keck_2021}, Subaru/SCExAO \citep{jovanovic_subaru_2015}, and SHARK-VIS \citep{pedichini_shark-vis_2022} and SHARK-NIR \citep{farinato_shark-nir_2022} at LBTI enable diffraction-limited imaging of circumstellar objects and study of spatial features at au scales. Instruments that can image at visible wavelengths like MagAO-X, SCExAO, and SHARK-VIS complement instruments that primarily work in the infrared like SPHERE, GPI, and KPIC to provide the means to probe the radiative transfer properties and structure of extrasolar dust across a significant part of the electromagnetic spectrum. However, despite significant gains in AO performance over the last decade, high-contrast images are still fundamentally-limited by speckle noise, leading to expected raw contrasts of about $10^{-4}$ (i.e., the ratio between the astrophysical signal to its host star; \citealt{guyon_limits_2005}). As such, there is value in the development in novel post-processing methods to increase contrast performance.

High-fidelity resolved images of circumstellar disks often prove to be a challenging feat because they are extended objects whose structure is not known \textit{a~priori}, complicating subtraction of the stellar point-spread function (PSF). The most commonly-employed observational strategies for high-contrast imaging observations are Angular Differential Imaging \citep[ADI;][]{marois_angular_2006} and Reference Differential Imaging \citep[RDI;][]{ruane_reference_2019}. When ADI or RDI is paired with advanced algorithms such as Principal Component Analysis (PCA) or Karhunen-Loéve Image Projection (KLIP; \citealt{soummer_detection_2012}), these methods have proven to be effective at reconstructing the majority of the structure associated with starlight which can then be subtracted from the data. However, one unfortunate tradeoff includes the unavoidable subtraction of the disk (or planet) signal along with the noise (\citealt{milli_impact_2012}; \citealt{pueyo_detection_2016}). To mitigate these problems, strategies based on the forward modeling (FM) of synthetic disk objects are being developed (e.g., \citealt{currie_resolving_2015};  \citealt{lawson_scexaocharis_2020}; \citealt{kueny_probing_2024}; \citealt{hom_uniform_2024}). In these techniques, the true detected disk signal is inferred by simulating the instrument response and PSF subtraction artifacts on a synthetic astrophysical object, often iteratively within a Markov chain Monte Carlo (MCMC) framework (\citealt{pueyo_detection_2016}; \citealt{mazoyer_diskfm_2020}). However, due to the large and complex parameter spaces associated with models of disks \citep{pueyo_detection_2016}, all possible morphologies of the disk must be explored for confidence in the final solution. This often leads to exhaustive modeling efforts that are computationally infeasible for all but the simplest of disk geometries.

In addition to these FM-based methods, several other post-processing methods have been proposed for better recovery of the uncorrupted disk signal in direct images. One such method is iterative PCA (IPCA), which builds upon the original KLIP/PCA algorithm by removing the current estimate of the disk object from the ADI dataset iteratively to get better estimates of the speckle field \citep{stapper_iterative_2022}. Recently, improvements to the RDI algorithm have been demonstrated using the star-hopping technique \citep{wahhaj_search_2021}, using an expansive archive of reference PSF images spanning several years \citep{xie_reference-star_2022}, and combining RDI with ADI as a new hybrid algorithm to exploit the strengths of both \citep{juillard_combining_2024}. Another method which has shown promise is data imputation through exclusion of the data containing the disk signal (\citealt{ren_using_2020}; \citealt{ren_karhunen-loeve_2023}). Finally, there has been considerable effort in recent years put towards joint estimations of the speckle field, disk/planet objects, and noise components through an inverse problem manner (REXPACO \citealt{flasseur_exoplanet_2018}; MAYONNAISE \citealt{pairet_mayonnaise_2021}; \texttt{mustard} \citealt{juillard_inverse-problem_2023}).

In this paper, we introduce a new, open-source Python pipeline \texttt{ffortissimo}, which we make freely available on GitHub\footnote{\url{https://github.com/jkueny/debrisdisk_freeform_fit_and_plot}}. \texttt{ffortissimo} uses a freeform, pixel-by-pixel, object retrieval methodology within a forward-modeling pipeline that is fully-differentiable. The rest of the paper is organized in the following manner: we provide additional background and a high-level overview of forward modeling in the context of KLIP-reduced images of disks in the next section. In Section \ref{sec:methods}, we provide details of our implementation of our pipeline and regularization methods. In Section \ref{sec:performance}, we report the performance observed thus far using on-sky MagAO-X data of the \hr{} disk in terms of the reconstruction of the intensity distribution of the disk signal through PSF subtraction artifacts. We discuss the remaining challenges that need to be addressed and limitations of our pipeline in Section \ref{sec:limitations}. Finally, we summarize our methods and results as well as identify avenues for future work in Section \ref{sec:summary}.

\section{Disk Forward-Modeling}
\label{sec:background}

\subsection{Rationale for a Data-Driven Disk Model}
\label{sec:rationale}
KLIP \citep{soummer_detection_2012} is a PSF subtraction algorithm that is essentially PCA, but assumes discrete-to-discrete operations with Gaussian statistics. These KLIP-induced side effects necessitate a careful balance, namely, tuning subtraction aggressiveness to maximize noise reduction while retaining the object's signal \citep{pueyo_detection_2016}; see \citealt{kueny_probing_2024}'s Figure 6 for a visualization using a disk surface brightness profile. As such, it is often the case that choosing an effective combination of KLIP reduction parameters is ambiguous. In the specific case of \adi{}, disks incur distortions because the disk is also present in the library of reference PSF images leading to self-subtraction.

To improve upon the abovementioned issues with fitting data using an assumed parametric form, one strategy is to instead utilize a bespoke model for fitting the dust density and brightness distributions. This bespoke model could perhaps leverage an ensemble(s) of basis functions, similar to what was demonstrated by \cite{han_recovering_2025} in 1D. However, fitting a full model requires detailed prior knowledge of the disk morphology of interest and may require a number of free parameters that would prove infeasible for iterative forward modeling in the style of, e.g., \cite{chen_multiband_2020}.


Detecting the scattering behavior across the entire azimuthal range of an inclined disk is of particular value when investigating exoplanet composition. However, recovering the minor axis of an inclined disk requires forward-modeling to recover the flux lost to PSF subtraction. Traditional forward modeling pipelines heavily rely on analytical approximations for the SPFs, most notably the Henyey-Greenstein (HG; \citealt{henyey_diffuse_1941}) parametrization. While convenient in the sense that a disk's SPF can be modeled by (typically) 1 to 3 parameters, such simplified functions enforce a shape unlike that of the complex scattering behaviors of realistic astrophysical dust \citep{hughes_debris_2018}. As a consequence, the model may produce qualitatively-good fits by leveraging degeneracies between SPF and geometrical parameters during model fitting (e.g., a disk model's stellocentric offsets can artificially brighten or dim parts of the disk; see \citealt{kueny_probing_2024}'s Figure 5 for a concrete example of this model degeneracy). A data-driven SPF reconstruction can capture complexities in the scattered starlight without a pre-supposed profile, allowing for a direct retrieval of the scattering properties of the dust. As simulations of the scattering due to realistic astrophysical dust continue to get more advanced \citep[e.g.,][]{lin_glitterin_2025}, an SPF inferred in this manner can help place robust constraints on dust grain composition and geometry without the need for the expensive forward modeling step.

Beyond scattering properties, a pixel-by-pixel framework unlocks the ability to probe morphological structures at or below the diffraction limit of the telescope. High-fidelity retrieval of sharp, localized features is critical for untangling the dynamical signposts of potential companions \citep[e.g.,][]{li_challenge_2024} or catastrophic collisional events \citep[e.g.,][]{kalas_second_2026}. Directly optimizing a high-resolution spatial grid allows these sub-resolution clumps, warps, and density discontinuities to emerge naturally from the data, providing a clearer window into the active dynamical environments of young planetary systems.

\subsection{Revisiting forward modeling with KLIP}

The full mathematical and array shape descriptions behind KLIP forward modeling (KLIP-FM) generalized for spectral cubes (i.e., data collected using an integral field spectrograph) can be found in \cite{pueyo_detection_2016}. Though the original description of KLIP-FM was developed in the context of imaging point sources, it can be applied to extended objects without much modification. We review the basic mathematical concepts behind KLIP-FM for disks (i.e., \textsc{DiskFM}; \citealt{mazoyer_diskfm_2020}) for single-filter imaging below.

With focal plane coordinates $\boldsymbol{x}$, we can express a given target image at a time of exposure $t$ as, 

\begin{equation}
T(\boldsymbol{x}) = S_{\psi_t}(\boldsymbol{x}) + \epsilon A(R_{\theta_t}[\boldsymbol{x}])
\end{equation}

where $S_{\psi_t}(x)$ is the speckle field intensity distribution at the focal plane, $\epsilon$ the photometry of the disk, and $A(R_{\theta_t}[\boldsymbol{x}])$ the image of the disk that is rotated according to the parallactic angle $\theta_t$ of the field-of-view through ADI. It is the speckle noise $S_{\psi_t}(x)$ that we seek to remove through KLIP. We assume that a KLIP basis $Z_{k}(\boldsymbol{x})$ comprised of a number of modes $k$ has been calculated using optimal parameters for the extended object of interest and saved to disk (see \citealt{soummer_detection_2012} for details on the calculation of $Z_{k}(\boldsymbol{x})$). One strategy for choosing the KLIP reduction parameters that maximize the $S/N$ of a disk is performing a grid search analysis over a range for each parameter for a given observation, as the optimal parameters are generally not known \textit{a priori} \citep{kueny_probing_2024}. For the specific case where the astrophysical source is also located in the library of reference PSFs (i.e., a typical ADI dataset), we express the resulting basis of KL modes $\{Y_k(\boldsymbol{x})\}_{k=1...K}$ as,

\begin{equation}
    Y_k(\boldsymbol{x}) = Z_{k}(\boldsymbol{x}) + \epsilon \Delta Z_k(\boldsymbol{x})
\end{equation}

where $\Delta Z_k(\boldsymbol{x})$ represents the perturbation of the KL basis by the disk object. Note for the RDI case where an astrophysical object is not present in the library of reference PSF images, $\epsilon = 0$. With this, the expression for subtracting the PSF estimate from an individual exposure $T(\boldsymbol{x})$ is:

\begin{equation}
    P_t(\boldsymbol{x}) = T_t(\boldsymbol{x}) - \sum_{k=1}^{K_{\text{KLIP}}} \langle T_t(\boldsymbol{x}), Y_k(\boldsymbol{x})\rangle_{\mathcal{S}} Y_k(\boldsymbol{x})
\end{equation}

where $P_t(\boldsymbol{x})$ denotes the PSF-subtracted image corresponding to exposure $t$, $\mathcal{S}$ represents the 2D search region within the field-of-view containing the object of interest, $K_{\text{KLIP}}$ represents the number of KL modes retained in the basis, and $\langle \cdot, \cdot\rangle_{\mathcal{S}}$ is our notation for the inner product over $\mathcal{S}$. In practice, to accelerate computation during disk analysis, the the search region $\mathcal{S}$ is typically an efficient support region matching the disk's expected morphology (see the leftmost thumbnail image in the ``Setup" lane in Figure \ref{fig:flowchart}). We can express the final KLIP-reduced image $P(\boldsymbol{x})$ as,

\begin{equation} \label{eq:derot}
    P(\boldsymbol{x}) = \sum_t P_t(R_{-\theta_t}[\boldsymbol{x}])
\end{equation}

where $R_{-\theta_t}[\boldsymbol{x}]$ denotes the image derotation operation according to the associated parallactic angle $\theta_t$. Equations through \ref{eq:derot} overview the basic operations concerning data reduction with KLIP on an ADI dataset to identify a faint disk.

As mentioned previously, one can forward model a synthetic disk object using the precomputed KL basis to calibrate against the artifacts induced by KLIP. In this case, the perturbed KL modes $\{\boldsymbol{\Delta Z}_{k}(\boldsymbol{x)}\}_{k=1...K}$ are computed using a disk model image $\widehat{A}(R_{\theta_t}[\boldsymbol{x}])$, copied and rotated by $\theta_t$ to generate an ensemble of models that mirrors the real disk signal present in an ADI dataset:

\begin{equation} \label{eq:perturb}
    \boldsymbol{\Delta Z}_{k}(\boldsymbol{x)} =\\
    \frac{\epsilon}{\sqrt{\Lambda_k}} \left(-\frac{1}{2\sqrt{\Lambda_k}} V_k^T \boldsymbol{\widehat{C}} V_k Z_k(\boldsymbol{x}) + V_k^T \boldsymbol{\widehat{A}_{\delta}}(\boldsymbol{x}) \right)
\end{equation}

where $\{\Lambda\}_{k=1...K}$ and $\{V\}_{k=1...K}$ are the eigenvalue and eigenvectors of the covariance matrix of the reference PSF library of $K$ images (available through the pre-computed KLIP basis), $\boldsymbol{\widehat{A_{\delta}}}(\boldsymbol{x})$ is a matrix populated with the rotated model disk images, and $\boldsymbol{\widehat{C}} = \boldsymbol{\widehat{A}_{\delta}S}^T + \boldsymbol{S}\boldsymbol{\widehat{A}_{\delta}}^T$ being the cross term between the concatenated reference PSF images $\boldsymbol{S}$ and the model images. We show a deconstructed version of Equation 66 from \cite{pueyo_detection_2016} to emphasize how the terms that estimate the phenomena of self- and over-subtraction are calculated in \textsc{DiskFM}:

\begin{equation}
    \boldsymbol{F}_{\text{over}}(\boldsymbol{x)} = \widehat{\boldsymbol{A}}_{\delta}(\boldsymbol{x}) - \sum_{k=1}^{K_{\text{KLIP}}} \langle \widehat{\boldsymbol{A}}_{\delta}(\boldsymbol{x}), Z_k(\boldsymbol{x})\rangle Z_k(\boldsymbol{x})
\end{equation}

and

\begin{equation}
    \begin{split}
        \boldsymbol{F}_{\text{self}}(\boldsymbol{x)} =  \sum_{k=1}^{K_{\text{KLIP}}} ( \langle \boldsymbol{T}(\boldsymbol{x}), Z_k(\boldsymbol{x})\rangle \boldsymbol{\Delta Z_k}(\boldsymbol{x}) \\
        + \langle \boldsymbol{T}(\boldsymbol{x}), \boldsymbol{\Delta Z_k}(\boldsymbol{x}) \rangle Z_k(\boldsymbol{x}) )
    \end{split}
\end{equation}

where $\boldsymbol{T}(\boldsymbol{x})$ is the target data matrix, $\boldsymbol{\widehat{A}}(\boldsymbol{x})$ is the matrix of synthetic disk object images, and $Z_k(\boldsymbol{x})$ represents the original KL modes accessed via the pre-computed KLIP basis. When combined, we show how a given disk object is forward modeled with the self- and over-subtraction effects due to KLIP applied:

\begin{equation} \label{eq:fm}
    \boldsymbol{F}(\boldsymbol{x}) = \widehat{\boldsymbol{A}}_{\delta}(\boldsymbol{x}) -  \boldsymbol{F}_{\text{over}}(\boldsymbol{x)} - \boldsymbol{F}_{\text{self}}(\boldsymbol{x)}.
\end{equation}

where $\boldsymbol{F}(\boldsymbol{x})$ is a matrix comprised of individual forward models $F_t(\boldsymbol{x})$. Recall, from Equation \ref{eq:perturb}, the term responsible for estimating the degree of self-subtraction scales as $\sim \epsilon / \sqrt{\Lambda_k}$ and so goes to zero in the RDI case. To wrap-up this high-level overview, we summarize the forward-modeling process in 3 steps:
\begin{enumerate}
    \item Create the processed image $P(\boldsymbol{x})$ using KLIP.
    \item Make a synthetic disk image $\widehat{A}(\boldsymbol{x})$ and model its movement through the dataset by applying the parallactic rotation consistent with the ADI observation $\widehat{A}(R_{\theta_t}[\boldsymbol{x}])$
    \item Use this model with Equation \ref{eq:fm} for every $Z_k(\boldsymbol{x}), V_k, \Lambda_k$ present in the saved KLIP basis to calculate the forward model for each science image $F_t(\boldsymbol{x})$, then derotate and collapse the cube of forward model images to obtain the final result: $F(\boldsymbol{x}) = \sum_t F_t(R_{-\theta_t}[\boldsymbol{x}])$
\end{enumerate}

With this, we can show the general forward-modeling cost function $\tilde{f}$,

\begin{equation}
    \tilde{f} = \text{arg } \underset{\widehat{f}}{\text{min }} ||P(\boldsymbol{x}) - \widehat{f}F(\boldsymbol{x}) ||_{\mathcal{S}}^2
\end{equation}

where $\widehat{f}$ is the estimated photometry of the disk object. Equations \ref{eq:perturb} through \ref{eq:fm} illustrate the main \textsc{DiskFM} workflow that is implemented iteratively to best estimate $\widehat{f}$. These calculations involve basic matrix operations that scale with the size and number of images in the image dataset as well as the number of modes retained in the KLIP basis. This, coupled with the need to exhaustively test the expansive parameter spaces associated with extended objects, illustrates the computational expense associated with \textsc{DiskFM}. However, recent gains in the capabilities and availability of graphics processing units (GPUs) have made projects involving expensive linear algebra operations more approachable. As such, by treating the pixels in a region of interest (ROI) as individual parameters when forward-modeling, one can use an optimization routine (e.g., gradient descent) to infer the true signal of the disk without needing to make strong assumptions on the disk's morphology and drastically cut down on computation time.

\begin{table*}[!ht]
\raggedright
\caption{MagAO-X Observation Log of \hr{} and PSF Reference}
\begin{tabular}{lclccccc}
\hline
Date        & Time Start/End (UT) & Object   & Band  & $t_{\text{exp}}$ (s) & $N_{\text{exp}}$ & $N_{\text{coadds}}$ & $\phi (^{\circ})$ \\ \hline
2023 Mar 10 & 05:49/07:36         & HR 4796A & $i'$  & 1                    & 5047             & 341                 & -21.5/65.0        \\
2023 Mar 10 & 05:52/07:36         & HR 4796A & $z'$  & 1                    & 4827             & 332                & -18.5/65.0         \\
2023 Mar 13 & 04:46/07:15         & HR 4796A & $r'$  & 0.23                 & 35355            & 465                & -58.5/61.5         \\
2024 Mar 29 & 06:44/06:52         & HR 4796A & $i'$  & 1                    & 2583             & 143                &  0.8/69.6          \\
2024 Mar 29 & 05:58/06:03         & HR 4748  & $z' $ & 1                    & 1540             &  83                & 31.4/75.6          \\
2024 Mar 29 & 05:22/05:31         & HR 4748  & $i'$  & 1                    & 1548             &  83                &  0.0/87.3          \\
2024 Mar 29 & 06:56/07:19         & HR 4796A & $z' $ & 1                    & 2683             & 141                & 8.0/69.6           \\
2025 Apr 09 & 03:29/05:47         & HR 4796A & $g'$  & 0.5                  & 15299            & 435                & -42.0/49.0         \\
\hline
            &                     &          &       &                      &                     &                
\end{tabular}

\footnotesize
\raggedright
\textbf{Notes.} $t_{\text{exp}}$: single frame exposure time; $N_{\text{exp}}$: number of raw science images; $N_{\text{coadds}}$: number of centered thumbnail images used for KLIP;  $\phi$: start/end parallactic angle. The total integration time for a given dataset is found by multiplying $t_{\text{exp}}$ and $N_{\text{exp}}$.
\label{tab:obslog}
\end{table*}

\section{Implementation Details}
\label{sec:methods}

The last decade's advances in deep learning \citep[e.g.,][]{lecun_deep_2015,goodfellowNIPS2016Tutorial2016a} have enabled powerful features in machine learning libraries such as PyTorch \citep{paszke_pytorch_2019}, TensorFlow \citep{abadi_tensorflow_2016}, and the framework we make use of in this paper, \textsc{Jax} \citep{bradbury_jax_2018}. \textsc{Jax} features an Application Programming Interface (API) resembling that of the popular Python library \textsc{NumPy} while enabling automatic differentiation (``autodiff"), Just-In-Time (JIT) compilation, and GPU-accelerated array computations. Since autodiff is the feature enabling the core of the capabilities of our pixel-by-pixel freeform modeling, we briefly touch on its description here. At a high level, and provided that the computations are carried out within the structure of the appropriate software library of choice (i.e., array operations are carried out using \textsc{Jax} functions and \textsc{Jax} arrays), autodiff facilitates the calculation of partial derivatives for numerical functions with respect to their floating point arguments by repeated execution of the chain rule of differential calculus. With information on the gradients, an optimizer is able to make more efficient decisions in regards to step size and direction across a complex parameter space as was demonstrated by \cite{millar-blanchaer_jwst_2025} through a \textsc{Jax}-based parametric disk modeling analysis.

\subsection{Data description}
\label{sec:data_description}

\begin{figure*}[!ht]
    \centering
    \includegraphics[width=\linewidth]{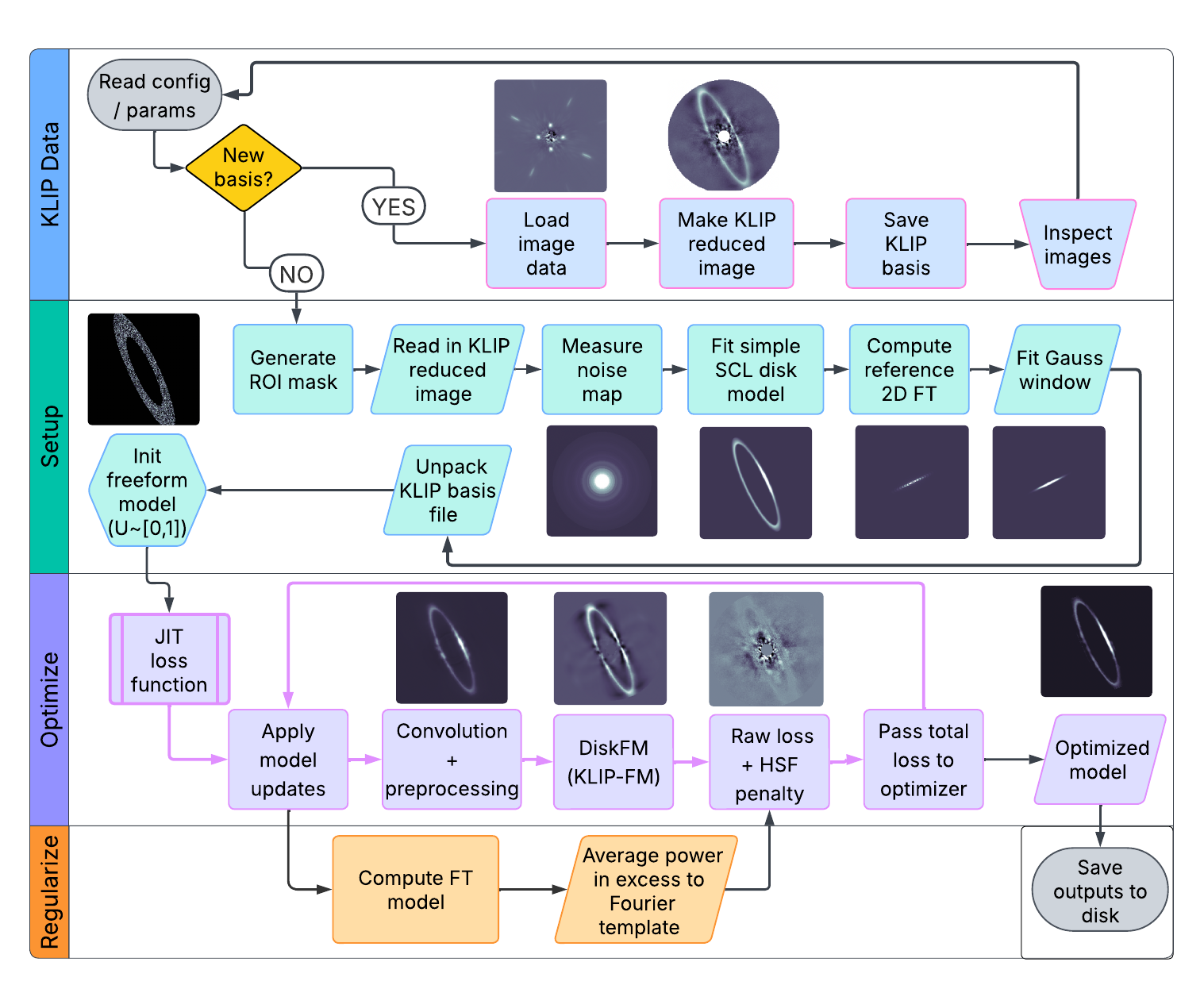}
    \caption{Flowchart illustrating the various processes and data used during the freeform model optimization. This workflow is explained in detail in Section \ref{sec:methods}. We inset small thumbnail images illustrating data products and the state of the freeform model during the main optimization loop next to the appropriate nodes in the diagram. We use standard ANSI flowchart shapes to denote processing steps, manual operations, data products, etc.}
    \label{fig:flowchart}
\end{figure*}

For this demonstration, we made use of on-sky image data of the iconic \hr{} disk which surrounds  a young A0 star of estimated age $8 \pm 2$ Myr \citep{stauffer_age_1995} located at a distance of 72.248 pc \citep{gaia_collaboration_gaia_2023}. \hr{} hosts a bright, highly inclined ring of dust that makes it ideally suited for benchmarking AO instrument performance and also for studying its grain properties because of the broad range of scattering angles that can be measured.

We collected these data using the ''extreme" Magellan Adaptive Optics instrument (MagAO-X) \citep{males_magao-x_2024}; further details on the observations and science results are described in Kueny et al. (2025, submitted). We use data from observations with the 6.5 m Magellan-Clay telescope at Las Campanas Observatory (LCO) during three separate epochs: 2023A, 2024A, and 2025A. For these observations, we chose to use ADI for the data taken in 2023A and 2025A and RDI for the data from 2024A using the star HR 4748 as a PSF reference; see Table \ref{tab:obslog}. Additionally, we made use of the small Lyot coronagraph (focal plane mask diameter of $3 \lambda/D$ at $656$ nm) for all observations. The instrument has a plate scale of 0\farcsec0059 \citep{long_astrometric_2025} and a field-of-view of $6\farcsec \times 6\farcsec$. However, to speed up computations, we spatially binned all images used in this demonstration with $2\times2$ binning for an effective plate scale of $0\farcsec012$. After every coronagraphic science observation, as part of collecting our calibration data, we removed the focal plane mask and collected about 5 minutes of unsaturated images of the star for convolution operations with our disk models.

\subsection{Image registration} \label{sec:registration_subsec}

We briefly recount the image processing steps for the image data before running KLIP; see Kueny et al. (2025, submitted) for a thorough description of all image processing steps described in this section. After excluding lower quality images, we prepared the raw images for PSF subtraction by subtracting a camera dark calibration frame and then normalizing the image pixel values by the exposure time. 

For image registration and snipping out our thumbnail images, we computed the approximate location of the star behind the coronagraph then used this coordinate to extract $224 \times 224$ ($2\farcsec69$ square) thumbnail images. To prepare for image registration, we made a bespoke software mask that isolated all artificial speckles (from the deformable mirror; DM; see \citealt{males_magao-x_2024} for an example MagAO-X PSF with these artifacts) in the field of view using elliptical apertures. To prepare the speckles in the stacked reference image as registration fiducials, we performed two pre-processing steps: (1) compute and subtract the median radial profile and apply an unsharp mask to remove the background and enhance the DM speckles, (2) apply our bespoke DM speckle mask to zero everything in the image besides these sharpened DM speckles. We repeated these exact pre-processing steps for every individual science thumbnail image. Then, we compute the needed x- and y-shift to center the PSF for every science thumbnail image by using phase cross-correlation \citep{guizar-sicairos_efficient_2008} on the isolated DM speckle image with the reference stacked image made by computing a median-stacked image using the entire set of (post-exclusions) thumbnails. Finally, we used these computed x- and y-shifts for every raw science thumbnail image. To reduce the computational burden during PSF subtraction and modeling procedures, we co-added our ADI images in time using parallactic angle binning of $0.25^{\circ}$ (shown under $N_{\text{coadds}}$ in Table \ref{tab:obslog}). For our RDI images, we co-added using 10 second bins since sky rotation is not as critical as in the ADI case \citep{wahhaj_search_2021}. Importantly, we chose these co-adding parameters because they ensure the disk does not smear more than the width of half of a pixel (5.9 mas) at the ansae (1060 mas from the star). This choice fully satisfies the Nyquist-Shannon sampling theorem at the disk's outermost points.

\begin{figure*}
    \centering
    \includegraphics[width=\linewidth]{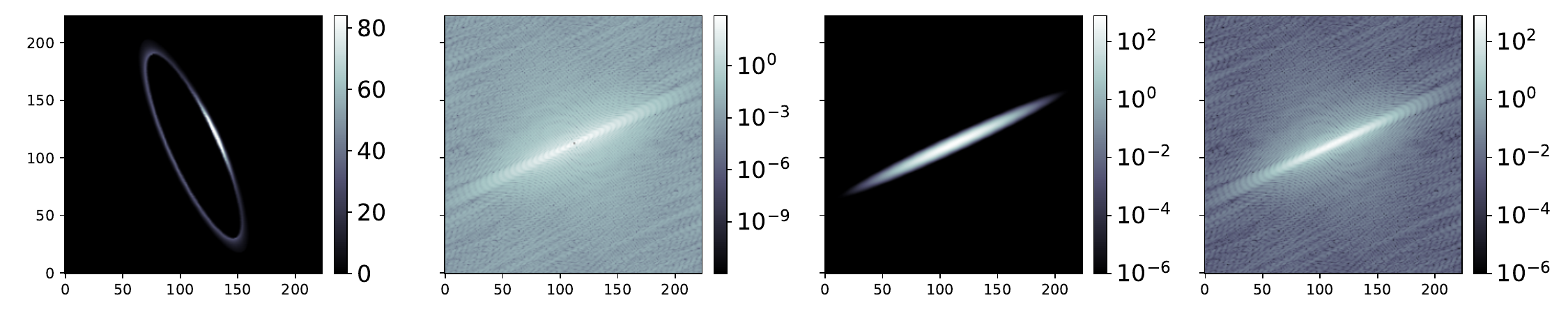}
    \caption{Suite of images illustrating how we create a reference template in Fourier space to regularize high-spatial frequencies during the freeform model optimization loop. \textbf{\textit{Left:}} a toy SCL disk model. \textbf{\textit{Middle left:}} the magnitude of the Fourier transform squared of the mean-subtracted toy disk model. \textbf{\textit{Middle right:}} the fitted elliptical Gaussian window function using the magnitude of the spatial frequency spectrum generously scaled by the maximum value of the magnitude spectrum image. \textbf{\textit{Right:}} the resulting spatial frequency template made by adding the fitted window function to the magnitude spectrum. We use this template in Fourier space to penalize freeform model spatial frequencies that are in excess to this template during the main optimization loop. All panels are plotted in log scale except for the toy model image, which is plotted linearly for clarity.}
    \label{fig:ref_freq}
\end{figure*}

\subsection{Data reduction with KLIP}
\label{sec:klip_subsec}

The routines associated with the data preparation phase of \texttt{ffortissimo} are illustrated in the top blue lane in Figure \ref{fig:flowchart} labeled ``Prep Data". The first step the code takes is reading in the configuration file which holds information about the various hyperparameters pertaining to the KLIP reduction, initial guesses for the disk model parameters, and file locations for the image data as well as metadata on the specific observation. Once the centered PSF image data are read in, the dataset is reduced using KLIP as implemented by \texttt{pyklip} \citep{wang_pyklip_2015}. The Karhunen-Loéve basis is saved for later use during the forward-modeling calculations to save considerable computational burden during the optimization loop. The steps detailed in this paragraph mirror those described in \cite{mazoyer_diskfm_2020} and in the original disk model fitting routine within an MCMC framework by J. Mazoyer.\footnote{\url{https://github.com/johanmazoyer/debrisdisk_mcmc_fit_and_plot.git}}

\subsection{Setup}
\label{sec:setup}

The processing to facilitate the regularization during the main freeform model optimization loop is illustrated in the ``Setup" lane in Figure \ref{fig:flowchart}. To start, we use the initial guesses for the disk's inclination and position angle contained in the configuration file to generate a binary software mask and define a disk-shaped ROI that is generously scaled spatially so as not to truncate any part of the disk. We then read in the KLIP-reduced image generated during the ``Prep Data" phase and apply the ROI mask. The masked KLIP-reduced image is then used to estimate the noise in concentric rings similar to \cite{kueny_probing_2024}.

The purpose of the following steps is to fit a crude disk model to our KLIP-reduced data, capturing the expected spatial frequencies in the disk model images. We then penalize high-spatial frequency content in the freeform disk model during the optimization. To obtain this reference disk model, we use the KLIP-reduced image to fit a simple scattered light (SCL) disk model using the L-BFGS \citep{liu_limited_1989} algorithm as implemented by the \texttt{scipy} library to perform gradient descent optimization over an arbitrary set of disk parameters. We made use of the disk model presented in \cite{ren_exo-kuiper_2019} which fits the dust density profile $\rho(r)$ with a radius of maximal dust density $r_c$ and two radial power laws ($\alpha_{in}$ and $\alpha_{out}$ to describe the dust density profile interior and exterior to $r_c$:

\begin{equation}
    \rho(r) \propto \left[\left(\frac{r}{r_c}\right)^{-2\alpha_{\text{in}}} + \left(\frac{r}{r_c} \right)^{-2\alpha_{\text{out}}} \right]^{-\frac{1}{2}},
\end{equation}

where $r$ denotes the radius from the star in au. The dust distribution in the vertical direction $Z(r,z)$ (perpendicular to the disk midplane) is modeled with a Gaussian profile,

\begin{equation}
    Z(r,z) \propto \exp{\left[-\left(\frac{z}{h_0r^{\beta}} \right)^2 \right]},
\end{equation}

where $z$ is the vertical distance, $h_0$ is the aspect ratio, and the exponent $\beta$ denotes the flaring parameter. For this analysis, we assumed no flaring is occurring, so $\beta = 1$ for all models we tested.

We generated the intensity $I$ for each pixel $(x', y')$ in the disk model using a brightness integral along the line of sight $z'$ (primed coordinates are with respect to the detector frame),

\begin{equation}
    I(x',y') = \int_{z' = -R_2}^{R_2} dz' \frac{N_0}{r^2}  \rho(r) Z(r,z) P(\theta),
\end{equation}

where $P(\theta)$ is the SPF, $R_2$ is the user-set outer disk radius where $I(x', y')$ defaults to zero, and $N_0$ the flux normalization factor. The disk is generated using a stellocentric coordinate system which then tilts with respect to the observer by the free parameter $\phi_{inc}$ and a second free parameter $\theta_{PA}$ adjusts the clocking of the model with respect to the sky. The scattering angle $\theta(x', y', z')$ is a function of the position of the image. We elected to use the two-parameter Henyey-Greenstein \citep[HG;][]{henyey_diffuse_1941} SPF which is a convenient mathematical function that approximates the forward- and back-scattering behavior seen in cosmic dust with only three free parameters. The functional form of the two parameter HG SPF is:

\begin{equation}
    \text{HG}_2(g_1,g_2,\alpha_1,\varphi) = \text{HG}(g_1,\varphi) + \alpha_1\text{HG}(g_2,\varphi)
\end{equation}

where $g_1$ and $g_2$ are the forward- and back-scattering asymmetry parameters respectively, $\varphi$ is the scattering angle, and $\alpha_1$ is a weighting factor. $g_1, g_2,$ and $\alpha_1$ are constrained to values between zero and one. HG is the SPF model and its functional form is:

\begin{equation}
    \text{HG}(g,\varphi) = \frac{1}{4\pi}\frac{1 - g^2}{{(1 - 2g\cos{\varphi}+g^2)}^{3/2}}.
\end{equation}

We note that we made use of this disk model for this specific disk and demonstration, but any other simple parametric disk model could be substituted for different geometries. Additionally, the reference disk model could be passed in by the user upon pipeline startup, foregoing the reference model generating procedure, if a model for the disk of interest already exists. For generating a quick reference model, we forgo forward modeling and only generate an image of the parametric disk model via convolution by the instrument PSF. This reference disk modeling procedure can be run on a personal laptop on the order of a few seconds; recall that this reference model only serves to mitigate non-physical features in the freeform disk model and can be roughly fit. We delve into the full rationale and validation process for using these data to regularize our freeform models in Section \ref{sec:regularization} below.

Once the optimizer converges to a reference \hr{}-like disk model $I_R(x', y')$, we calculate the modulus-squared of the Fourier transform (i.e., the power spectrum) to get the expected features in frequency space for this reference disk model:

\begin{equation} \label{eq:ref_spec}
    {\hat{\mathcal{R}}}(\xi, \eta) = \left| \mathcal{F}\left[ I_R(x', y') \right] \right|^2,
\end{equation}

where $\mathcal{F}$ denotes the Fourier transform operator to obtain a representation of $I_R(x', y')$ in frequency space. To ensure the optimizer concentrates on features of interest, we need to govern the model's spatial frequency content. We do this by employing another gradient descent optimization procedure to fit an elliptical Gaussian window function to the dominant feature in the reference spatial frequency spectrum. We apply a generous scale factor to the array element values of this fitted window function then add it to the reference frequency spectrum array to ensure that lower spatial frequencies are very unlikely to be penalized during regularization. Figure \ref{fig:ref_freq} illustrates this process with an example \hr{}-like toy disk model in the leftmost panel, the resulting magnitude spectrum of this toy disk model in the next panel, and the optimal elliptical Gaussian window function fit to this reference magnitude spectrum in the middle-right panel. We obtain the result shown in the rightmost panel by adding the fitted window function to the reference magnitude spectrum (i.e., adding the middle panels together). It is this template that is used to penalize excessively-high spatial frequencies during freeform disk model optimization.

We note that while we observed qualitatively-good model fits with just the spatial frequencies that the reference disk model is comprised of, we perform the addition of the fitted elliptical Gaussian window function to this spatial frequency spectrum as a guard against penalties toward low spatial frequencies. This was mainly motivated by the idea that extended diffuse emission may be penalized by our regularization method, as the reference disk models automatically generated by \texttt{ffortissimo} do not capture a component like this with our current data (see \citealt{schneider_hr_2018} to see the extended emission by small dust grains around the \hr{} system). However, more testing is needed to determine if an elliptical Gaussian is the ideal window function to fit for additional disk morphologies.

\subsection{Optimization}
\label{sec:optimization}

For optimizing the freeform disk model, we make use of the Adam \citep{kingma_adam_2014} algorithm implemented by the \texttt{optax} library \citep{deepmind_deepmind_2020} due to the complexity and number of free parameters in our modeling (recall, we are fitting every pixel in the modeling ROI as parameters). We show the processes and data involved in the main optimization loop in Figure \ref{fig:flowchart} in the ``Optimization" lane. We encapsulate the modeling and loss calculation routines in JIT to dramatically increase loop speed. At the beginning of each iteration, the model parameter updates are applied to the vector of free parameters (pixels) $\boldsymbol{\theta}$:

\begin{equation}
    \boldsymbol{\theta} = [\theta_1, \theta_2, \cdots, \theta_N] \in \mathbb{R}^{N}
\end{equation}

where $N$ constitutes the number of free parameters and is dependent on the size of the ROI defined during the setup phase. Since we are fitting for the intensity distribution of the disk, we enforce a positivity constraint before proceeding to avoid non-physical results. The model parameters $\boldsymbol{\theta}$ make up a subset of the flattened model vector $\boldsymbol{M}(x; \boldsymbol{\theta})$ where $M(x_i) = \boldsymbol{\theta_i}$ if $x_i \in  \text{ROI}$ for pixel $p_i$. The model vector is reshaped to a 2D array $(\boldsymbol{M} \rightarrow M(x', y';\boldsymbol{\theta}))$ to facilitate the convolution and image rotation operations down the line. Next, a copy of the current model is created, mean-subtracted, and normalized for regularizing high-spatial frequency content in the freeform model (see Section \ref{sec:regularization} below). A penalty for high-spatial frequency is calculated by first computing the power spectrum of the model in the current iteration:

\begin{equation}
    \hat{F}(\xi, \eta) = \left| \mathcal{F}\left[ C(x', y';\boldsymbol{\theta}) \right] \right|^2
\end{equation}

where $C(x', y';\boldsymbol{\theta})$ is the normalized copy of the current iteration's freeform model array. The magnitude of the spatial frequency spectrum of the current freeform model $\hat{F}(\boldsymbol{\theta})$ is compared to the reference spatial frequency template to compute the spatial frequency penalty value:

\begin{equation} \label{eq:penalty}
    \lambda_{\text{reg}} = \frac{1}{N_{\xi}} \sum_{ab} \begin{cases}
                                          \hat{F}_{ab}(\xi, \eta) - \hat{\mathcal{R}}_{ab}(\xi, \eta), & \hat{F}_{ab} > \hat{\mathcal{R}}_{ab} \\
                                          0, & \text{otherwise}
                                      \end{cases}
\end{equation}

where $N_{\xi}$ is the number of elements in the magnitude spectrum array $\hat{F}(\boldsymbol{\theta})$, $ab$ denote array indices, and $\hat{\mathcal{R}}(\boldsymbol{\theta})$ is the reference spatial frequency template created during the setup phase; recall, $\hat{F}(\boldsymbol{\theta})$ and $\hat{\mathcal{R}}(\boldsymbol{\theta})$ denote the magnitude of the spatial frequency spectrum of disk models so they only contain positive values. Note that because we built the entire forward-modeling procedure within the framework of \textsc{Jax}, piecewise conditional functions are intrinsically differentiable.

After the penalty is computed, the freeform model is convolved with the instrument PSF and optional preprocessing steps are applied if enabled in the configuration file. We apply preprocessing steps to the model image such as a radial profile subtraction and/or high-pass filtering to mirror the steps applied to the raw image data when preparing for PSF subtraction (see Section \ref{sec:klip_subsec}). The forward model is calculated using the routines employed by the \textsc{DiskFM} module which we converted into \textsc{Jax} operations.

After the forward model is calculated, it is compared to the KLIP-reduced image by calculating the square of the residuals normalized by the empirically-measured noise map:

\begin{equation}
    R(x', y';\boldsymbol{\theta}) = 
 \frac{[D(x', y') - M(x',y';\boldsymbol{\theta})]^2}{\sigma(x', y')}
\end{equation}

where $D(x',y')$ is the KLIP-reduced image, $M(x',y';\boldsymbol{\theta})$ is the forward model, and $\sigma(x', y')$ is the empirically-measured spatial noise map. We note that during initial development, we experimented with feeding the standard loss $R(x', y';\boldsymbol{\theta})$ into the Huber Loss \citep{huber_robust_1964} which is better suited at fitting disk features in the presence of speckle noise, which is non-Gaussian \citep{fitzgerald_speckle_2006}. However, fitting the \hr{} disk proved difficult; the optimizer treated the high signal-to-noise outermost regions as outliers compared to the fainter minor axis. Most importantly, the signal from the outermost portions of the disk are not speckle noise-limited. We plan to study alternative strategies for implementing the Huber Loss in the loss calculation in a future paper.

The high-spatial frequency penalty is added to $R(x', y';\boldsymbol{\theta})$ to compute the total loss $L_T(\boldsymbol{\theta})$:


\begin{equation} \label{eq:loss}
    L_T(\boldsymbol{\theta}) = H_{\delta}(\boldsymbol{\theta}) + \lambda_{\text{reg}}
\end{equation}

where $\lambda_{\text{reg}}$ is defined in Equation \ref{eq:penalty}. $L_T(\boldsymbol{\theta})$ is passed into the optimizer that decides the model updates to apply for the next iteration to lower it. These optimization steps iterate until the loss value plateaus. Convergence is usually reached after around 10k iterations taking around 1 to 2 hours of run time on a NVIDIA H100 GPU.

\begin{figure}
    \centering
    \includegraphics[width=0.5\linewidth]{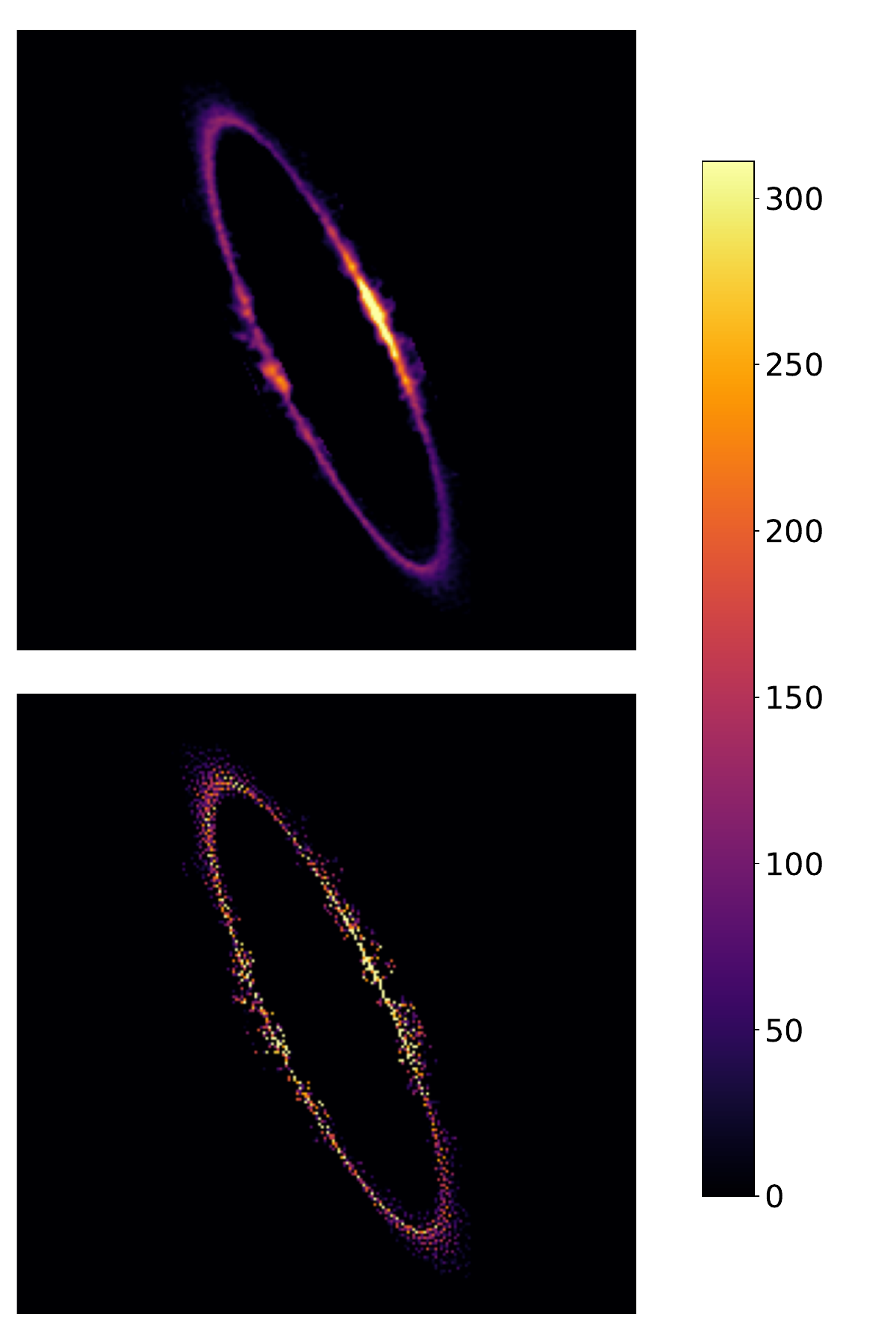}
    \caption{Showcase of the effects of spatial frequency regularization. \textbf{\textit{Top:}} Example optimized freeform disk model with regularization. \textbf{\textit{Bottom:}} Optimized model using the same dataset and modeling parameters as the model in the top panel but without regularization. The colorbar represents pixel counts in arbitrary units.}
    \label{fig:regularization}
\end{figure}

Preliminary modeling runs revealed that regularization is a necessary and important aspect of our freeform modeling pipeline. When regularization is not used, we found that the freeform disk model overfitted the noise and converged to a model that appeared ``gritty" and only appeared as a qualitatively good fit to the disk after convolution with the instrument PSF; see Figure \ref{fig:regularization} for a comparison between an example regularized disk model versus one that was optimized without regularization. We mentioned the steps taken to regularize high-spatial frequency content in Section \ref{sec:setup} above. We describe our efforts towards validating our use of a reference spatial frequency spectrum below.

\subsection{Regularization validation testing}
\label{sec:regularization}

\begin{figure*}[ht!]
    \centering
    \includegraphics[width=0.9\linewidth]{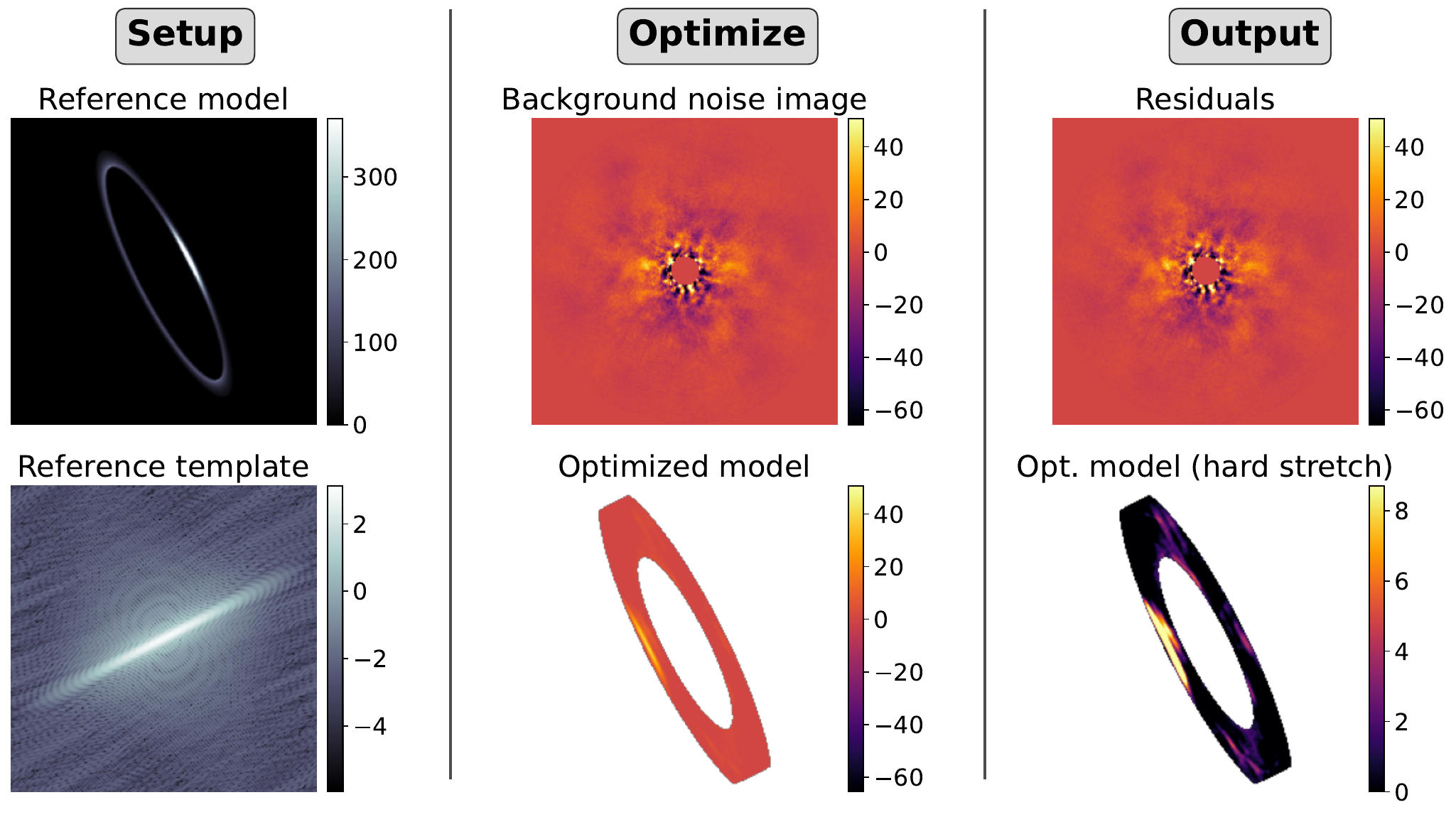}
    \caption{Results of the test we performed to probe model biases with our method for regularizing high-spatial frequency content in the freeform model. \textbf{\textit{First column:}} the top image shows the reference disk model. We show the reference power spectrum template in the bottom panel. \textbf{\textit{Second column:}} the top image shows the diskless KLIP-reduced image that we fit a freeform model to using the same setup parameters we would use when fitting a real disk image. The bottom panel shows the optimized freeform model fit to the diskless image after 10k iterations. \textbf{\textit{Third column:}} The top panel shows the residuals between the diskless image and the optimized freeform forward model and the bottom panel again shows the optimized freeform model, but this time with a harder colorbar stretch. The result of this test suggests that our regularization methods are not biasing our freeform models in a significant way.}
    \label{fig:bkgfit}
\end{figure*}

We employ the abovementioned automatic fitting procedure using a toy disk model to the KLIP-reduced image data as a low-complexity prior for generating the reference spectrum $\mathcal{I}(\xi,\eta)$, as shown in Equation \ref{eq:ref_spec}. Equation \ref{eq:penalty} shows how we use the average of the power in excess to our spatial frequency template $\hat{\mathcal{R}}_{ab}(\boldsymbol{\theta})$ to increase the loss in Equation \ref{eq:loss}. We emphasize that this regularization method is a way to enforce object smoothness, physicality, and sparsity all with a single regularization parameter, dramatically simplifying the work needed to learn the behavior of the regularization between datasets and objects. 

This method can perhaps be compared with matched filtering which is widely used in the field of signal processing. In short, a matched filter is the linear filter or template that maximizes the signal-to-noise of a known signal and has proven to be an effective method for finding exoplanets in direct images (\citealt{cantalloube_direct_2015}; \citealt{ruffio_improving_2017};\citealt{de_rosa_direct_2023}). In our current methods, we create a template in Fourier space of the expected spatial frequencies for a realistic disk of similar morphology to the disk being characterized. During every iteration in the main optimization loop, we evaluate the spatial frequency spectrum of the current freeform model and penalize excessive spatial frequency content using our reference template. In this way, we encourage a freeform model that is smooth but is still permitted to have edge profiles consistent with a realistic disk.

\begin{figure*}[ht!]
    \centering
    \includegraphics[width=0.9\linewidth]{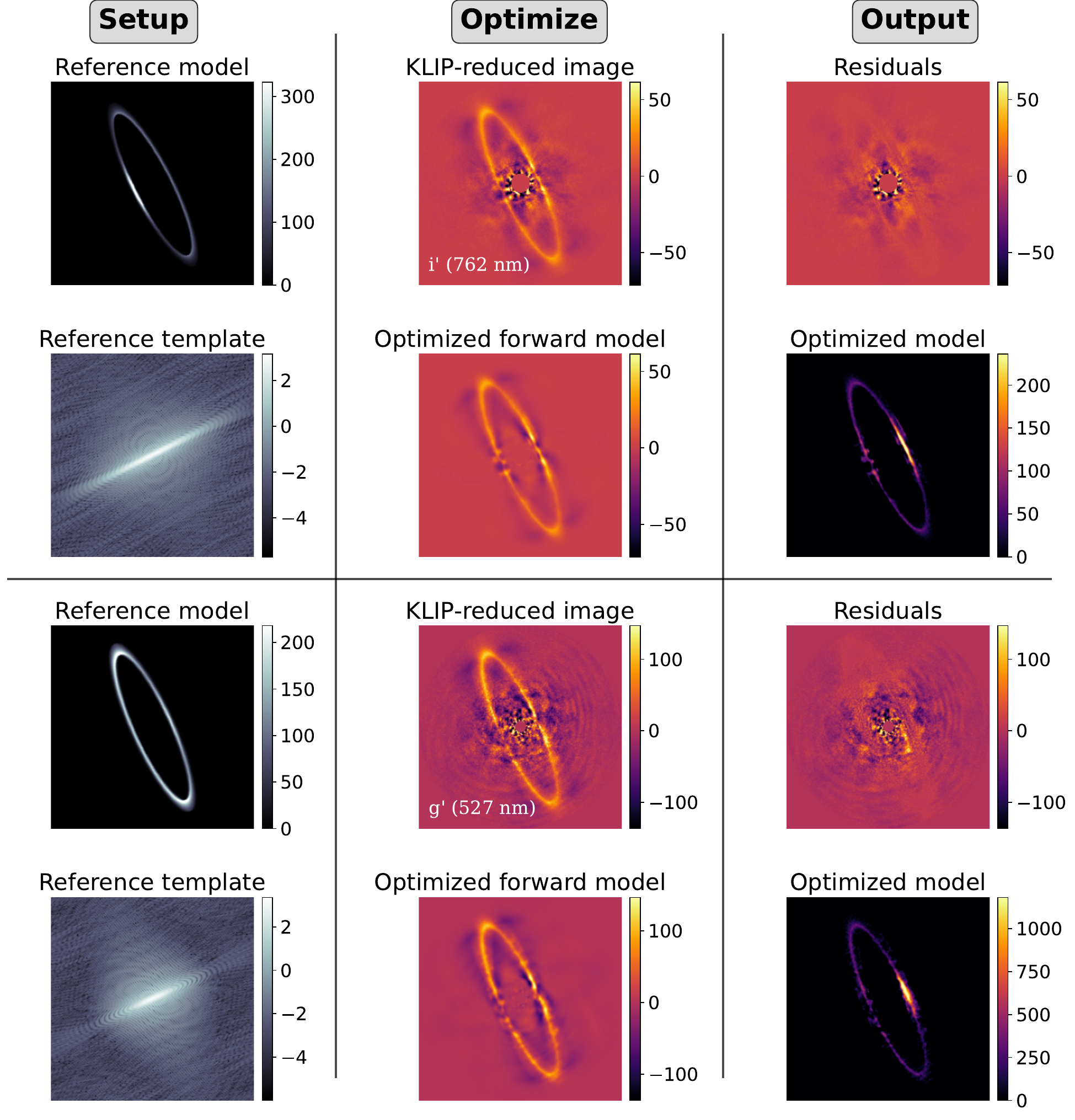}
    \caption{Showcase of two experiments to validate the use of simple reference disk models for regularizing the freeform model. The top 6 panels show results when using a reference disk model with a bright spot in a distinctly different location than the expected location when optimizing the \iband{} dataset from UT 2023-03-10. The bottom 6 panels show that using a disk model that is nearly isotropic in brightness still produces the expected morphology when fitting a model to the \gband{} image from UT 2025-04-09. \textbf{\textit{First column:}} the top image shows the reference disk model that we used to calculate the magnitude spectrum of the expected spatial frequencies of a real disk which use as a template, shown in the bottom panel. \textbf{\textit{Second column:}} the top image shows the KLIP-reduced image that we fit a freeform model to using the same setup parameters we would use when fitting a real disk image. The bottom panel shows the optimized freeform model after 10k iterations. \textbf{\textit{Third column:}} the top panel shows the residuals between the diskless image and the optimized freeform forward model and the bottom panel shows the optimized freeform model. These test results are good evidence that our regularization methods are not biasing our freeform models in a significant way.}
    \label{fig:experiments}
\end{figure*}

To probe whether our regularization methods were biasing our freeform models, we performed three tests utilizing different data sets. In the first test, we probed for model biases by fitting a freeform model to a ``diskless" KLIP-reduced image using the \iband{} \adi{} dataset which we illustrate in Figure \ref{fig:bkgfit}. For the first step during this test, we generated a KLIP-reduced image of just background noise by performing KLIP but ``back-rotating" using the negative of the parallactic angles for each PSF image as shown in \citet{gerard_planet_2016} causing the disk to ``median out" of the image. We then fit a freeform model to the diskless image using the same methods as before. Concretely, we generated a reference toy disk model using the best-fit disk model parameters from \cite{chen_multiband_2020} (we did not use the normal crudely-fit disk model for a more robust test) and made the reference disk frequency template from this idealized reference disk. Additionally, we increased the scale factor applied to the spatial frequency penalty $\lambda_{\text{reg}}$ by two orders of magnitude to force the optimizer to make use of the frequencies contained almost entirely within the dominant diagonal-like feature in the reference disk frequency spectrum (see Figure \ref{fig:bkgfit}, bottom left image). The results of this test showed that our regularization methods are unlikely to introduce biases in the freeform model as there was no semblance of disk-shaped features in the optimized freeform model fit to the diskless image.

We validated our methods further with two additional tests, where we used reference disk models that were (1) flipped along the major axis or (2) generated without a bright forward-scattering peak as shown in Figure \ref{fig:experiments}. The purpose of these tests was to check if the prominent bright spot at the reference disk model's minor axis was biasing the forward-scattering peak in the freeform disk models. For the test involving the flipped major axis, we used the best-fitting disk model parameters from \cite{chen_multiband_2020} to generate a reference disk frequency template in Fourier space. For the test involving the absence of the forward-scattering peak we specifically chose the \gband{} dataset from UT 2025 Apr 9 because the minor axis is so heavily self-subtracted due to \adi{} that our quick reference disk model fitting procedure produces an essentially isotropic reference disk model. As shown in the ``Output" column in Figure \ref{fig:experiments}, even though the reference models presented a distinctly-different brightness distribution than the expected result, the optimized models still led to good residuals with the KLIP-reduced images. The results of these tests prove that the bright minor axis feature of the reference disk model does not meaningfully influence an identical feature in the freeform models.

\section{Performance}
\label{sec:performance}

To evaluate the utility of our pipeline, we benchmarked its ability to characterize complex disk morphology and its fidelity in extracting SPFs. We subjected our \hr{} datasets to a separate forward-modeling analysis using a traditional SCL disk modeling approach using \textsc{DiskFM} wrapped in an MCMC framework (see Kueny et al. 2025, in review). Discussing an apples-to-apples gain in computation speed is complicated because of the markedly different approaches (e.g., a parallelized MCMC calculation versus GPU-accelerated gradient descent optimization), but in general we observed speedups of a factor of roughly 100---150X for the freeform modeling over the MCMC \textsc{DiskFM} approach. For our MCMC \textsc{DiskFM} analysis, we made use of a 94-core node in a computing cluster that ran for several days of wall clock time to reach the convergence criteria. Conversely, running \texttt{ffortissimo} with our largest datasets on a single NVIDIA H100 GPU takes $\sim 1.5$ hours to converge. Due to the gains in computation speed, we were able to double the amount of images in our datasets by co-adding fewer raw images, which also increased the signal-to-noise in the final reduced images.

As \texttt{ffortissimo} is based on \textsc{DiskFM} \citep{mazoyer_diskfm_2020}, it currently supports \adi{} and \rdi{} reductions; work to increase support for a more diverse spread of PSF subtraction algorithms is ongoing. So far, we have tested our pipeline on MagAO-X data, however, data taken with other AO instruments should be compatible along with modifications to the standard configuration file and provided the data are not within a spectral cube format.

\subsection{Characterizing disk morphology}

Our pipeline results have shown that this implementation of a freeform model is an effective tool for morphological characterization using on-sky images of disks. SCL disk models generated using standard parametric descriptions \citep[e.g.,][]{augereau_hr_1999} often fail to produce qualitatively-good residuals when fit to disk images (see, e.g., \citealt{hom_uniform_2024}, \citealt{kueny_probing_2024}, or \citealt{facchini_2_2026}). As the number of known circumstellar disks increases and the capabilities of today's direct imaging instruments continue to improve, we are learning not only of the existence of geometrically complex features in extrasolar dust, but also that they are commonplace in disks (e.g., HD~106906; \citealt{kalas_direct_2015}; HD~141569A; \citealt{perrot_discovery_2016}; HD~142527; \citealt{li_challenge_2024}). Fine spatial features such as tight gaps or sharp warps make parametric disk models increasingly difficult to implement, complicating further study into these interesting features. While our freeform models do not explicitly provide physical constraints on disk features, they offer a representation of the disk dust in the absence of KLIP artifacts and blurring due to diffraction. These freeform models also serve as a vector to which one could fit another physically-motivated model to learn about the retrieved features directly. Ultimately, there is value in the study of complex dusty features in disks as it may inform the processes that drive planet formation or hint to the locations of unresolved planets.

\begin{figure}
    \centering
    \includegraphics[width=0.5\linewidth]{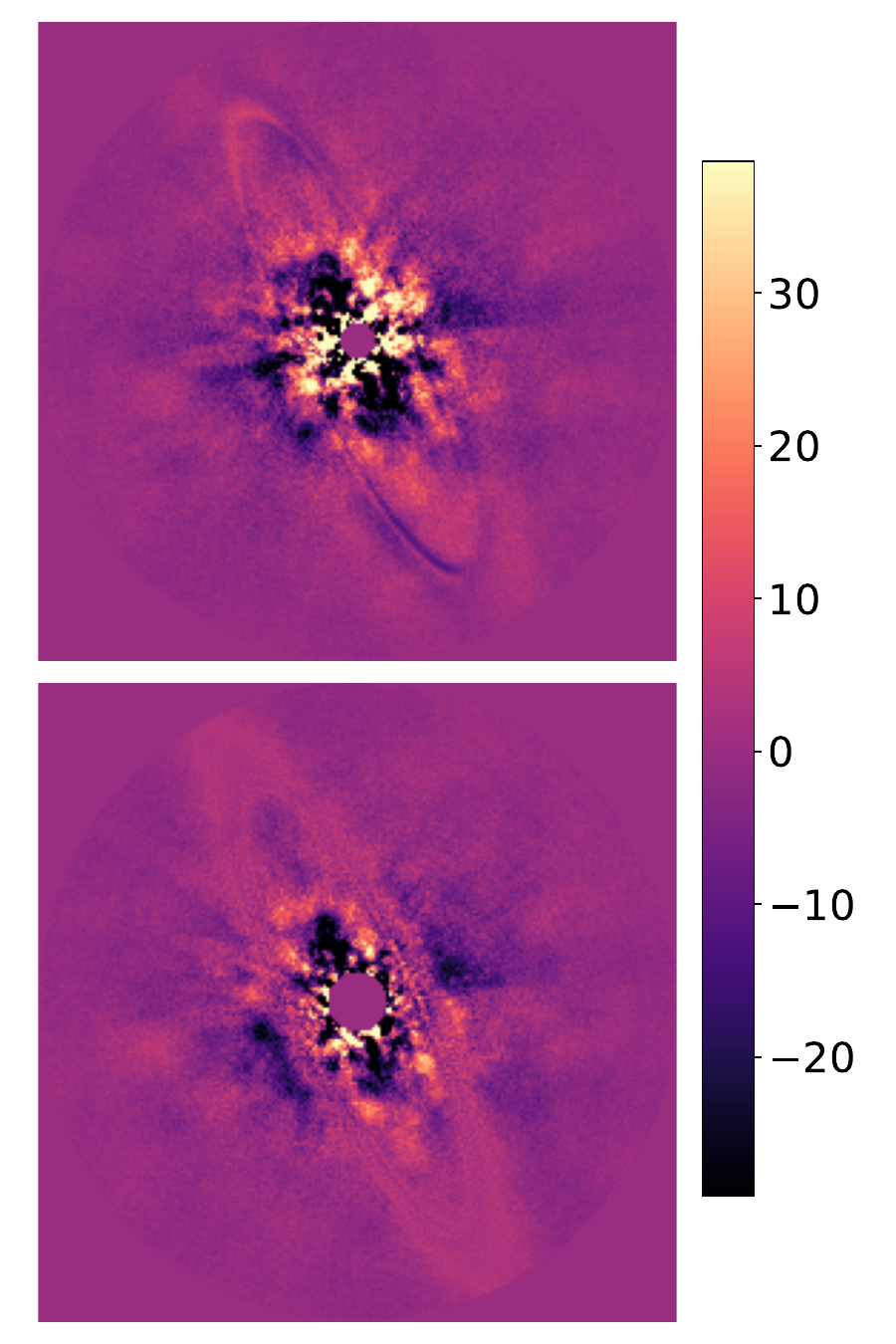}
    \caption{Comparison of residuals images. \textbf{\textit{Top:}} Residuals between the \adi{} disk image at \rband{} made with a traditional SCL disk model (as described in \citealt{ren_exo-kuiper_2019}). \textbf{\textit{Bottom:}} Residuals using the same \adi{} image but using an optimized freeform disk model. Though the oversubtraction artifacts and residual disk halo are not present in the freeform disk model residuals image, the freeform model learns some of the noise especially around the minor axis of the disk which is one of the remaining challenges associated with our pipeline.}
    \label{fig:residuals_comparison}
\end{figure}

\begin{figure*}[ht!]
    \centering
    \includegraphics[width=0.9\linewidth]{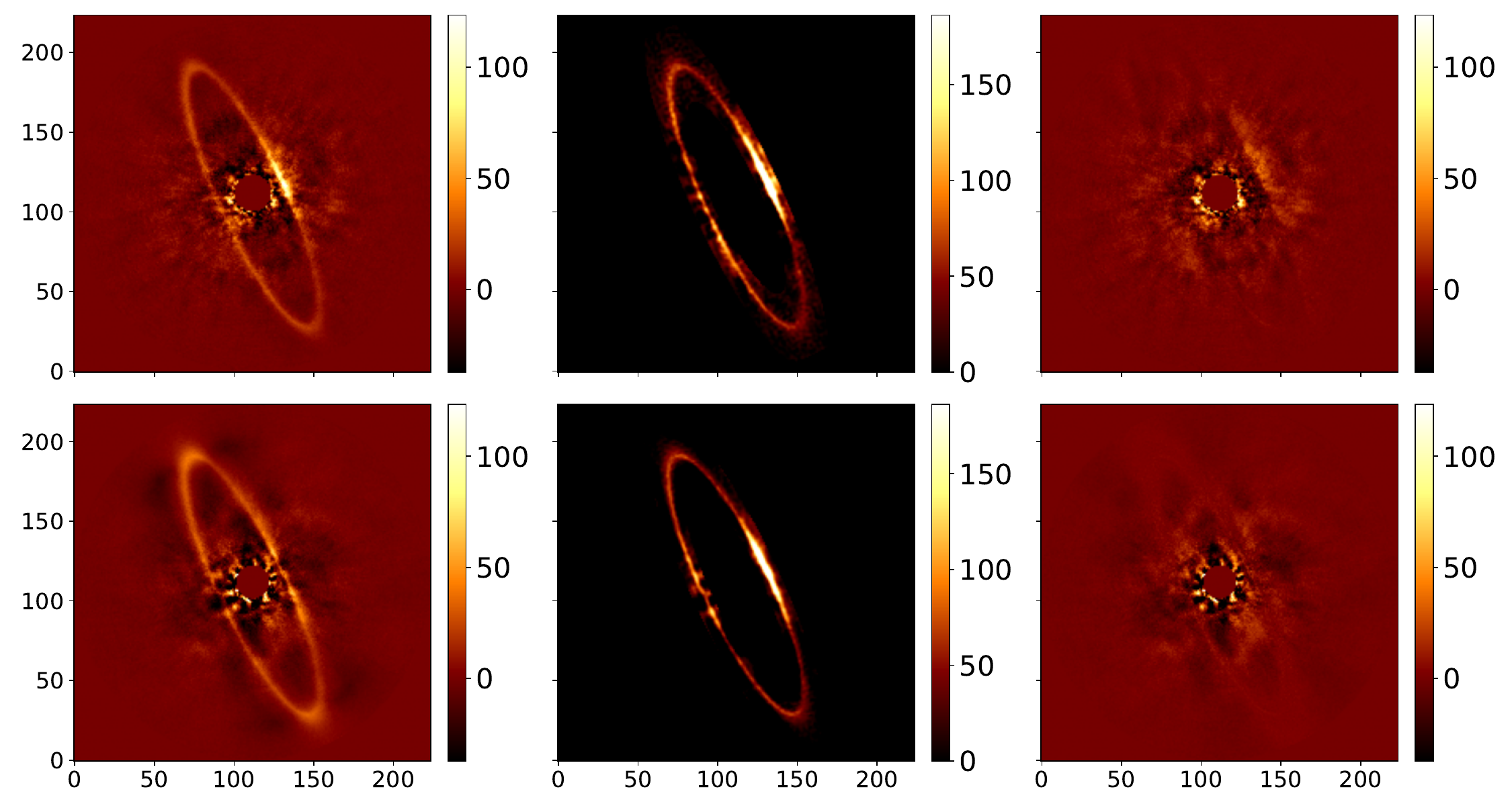}
    \caption{Comparison of freeform modeling results between \adi{} and \rdi{} datasets, illustrating inference of the disk through self-subtraction artifacts. \textbf{\textit{Top row:}} results for the \rdi{} image. The left panel shows the KLIP-reduced image of the \hr{} disk, the middle panel shows the optimized freeform model after 10k iterations, and the right panel shows the residuals between the KLIP-reduced data and the optimized freeform forward model. \textbf{\textit{Bottom row:}} The same layout as the top row, but for the case of the \adi{} image.}
    \label{fig:klip_comparison}
\end{figure*}

We subjected our \hr{} datasets to a previous forward modeling analysis using the parametric disk model description from \cite{ren_exo-kuiper_2019}, but we observed significant residuals despite extensive effort (see Kueny et al. 2025, submitted). We then subjected our data to our novel freeform modeling pipeline which showed improvement in the residuals images, especially because of the brightness asymmetry in the lobes (Figure \ref{fig:residuals_comparison}). However, one remaining challenge to address is that the freeform models learn noise from the observations especially at small angular separations where speckle noise is highest. However, a more persistent noise artifact that is learned by our freeform models is the wind-driven halo (WDH) which presents as a butterfly-shaped artifact spanning the AO dark hole region (see \citet{cantalloube_wind-driven_2020} for a clear example image of the WDH). Extended noise features due to the WDH are also seen in Figure \ref{fig:residuals_comparison} as the bright and dark bands extending radially-outward from the center of the image.

\subsection{Extracting Disk Scattering Phase Functions} \label{sec:spf}

Studying a disk's SPF allows inference of the dust composition and physical characteristics, which ultimately informs exoplanetary composition. However, since extrasolar dust grains are highly forward-scattering \citep{hughes_debris_2018}, it is necessary to probe a wide range of scattering angles including the forward and backward scattering peaks that like along the minor axis. The challenge lies in the fact that the minor axis for highly-inclined disks (which offer the greatest scattering angle coverage) is in a high speckle noise region (see, e.g., \citealt{hinkley_temporal_2007}; \citealt{males_mysterious_2021} for detailed descriptions of  speckle noise). Current practices for mitigating this type of noise at close inner-working angles (IWAs) are based heavily on the \adi{} algorithm. For the \hr{} disk specifically, the \adi{} algorithm almost completely erases the dramatic forward-scattering at the minor axis (Figure \ref{fig:klip_comparison}). As shown in the bottom right panel in Figure \ref{fig:klip_comparison}, our freeform forward modeling pipeline is one way to recover the disk signal lost by the \adi{} algorithm without a fixed-form parametric disk model.

\begin{figure*}[ht!]
    \centering
    \includegraphics[width=\linewidth]{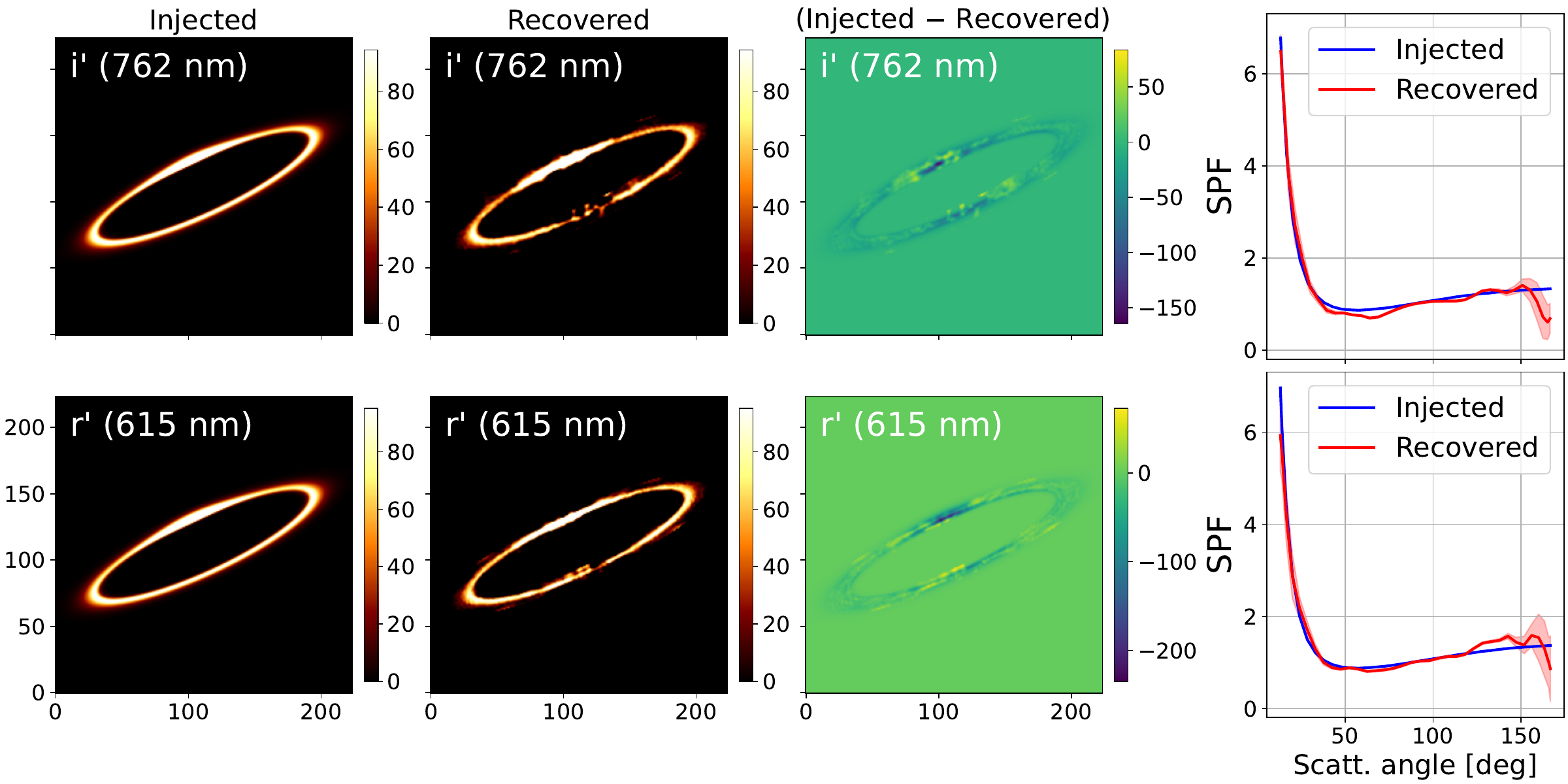} \\
    \caption{Inject-and-recover test results using a synthetic disk with a known SPF showing the degree to which \texttt{ffortissimo} can infer an SPF given different noise profiles. We used two different ``back-rotated" \adi{} datasets (see Section \ref{sec:regularization}) from separate nights. The top row shows the results using the \iband{} \adi{} data and the bottom row shows results using the \rband{} \adi{} dataset. For each row, in left to right progression, we display the synthetic disk model with a known SPF that was injected into the science images, the recovered freeform model after optimization, the difference image between the injected and recovered models, and the comparison of the injected vs. the recovered SPFs.}
    \label{fig:inject_recover}
\end{figure*}
    
For comparison, we fitted freeform models on our \adi{} and \rdi{} reduced images at \iband{} and \zband{} and then extracted the SPFs from those freeform models. We compare our extracted freeform SPFs with others measured using a parametric SCL disk model and custom SPF formed by a subset of the Legendre polynomials, similar to what was demonstrated in \cite{arriaga_multiband_2020}. Briefly, this comparison SPF extraction method fits the geometric parameters of a disk model in the usual sense, but uses the coefficients of the first $\sim 18$ Legendre polynomials to model the azimuthal brightness distribution of the disk within a Markov chain Monte Carlo framework; this Legendre polynomial-based SPF modeling technique is described in detail in \cite{kuenyMultibandStudyHR2026}.

\begin{figure*}
    \centering
    \includegraphics[width=\linewidth]{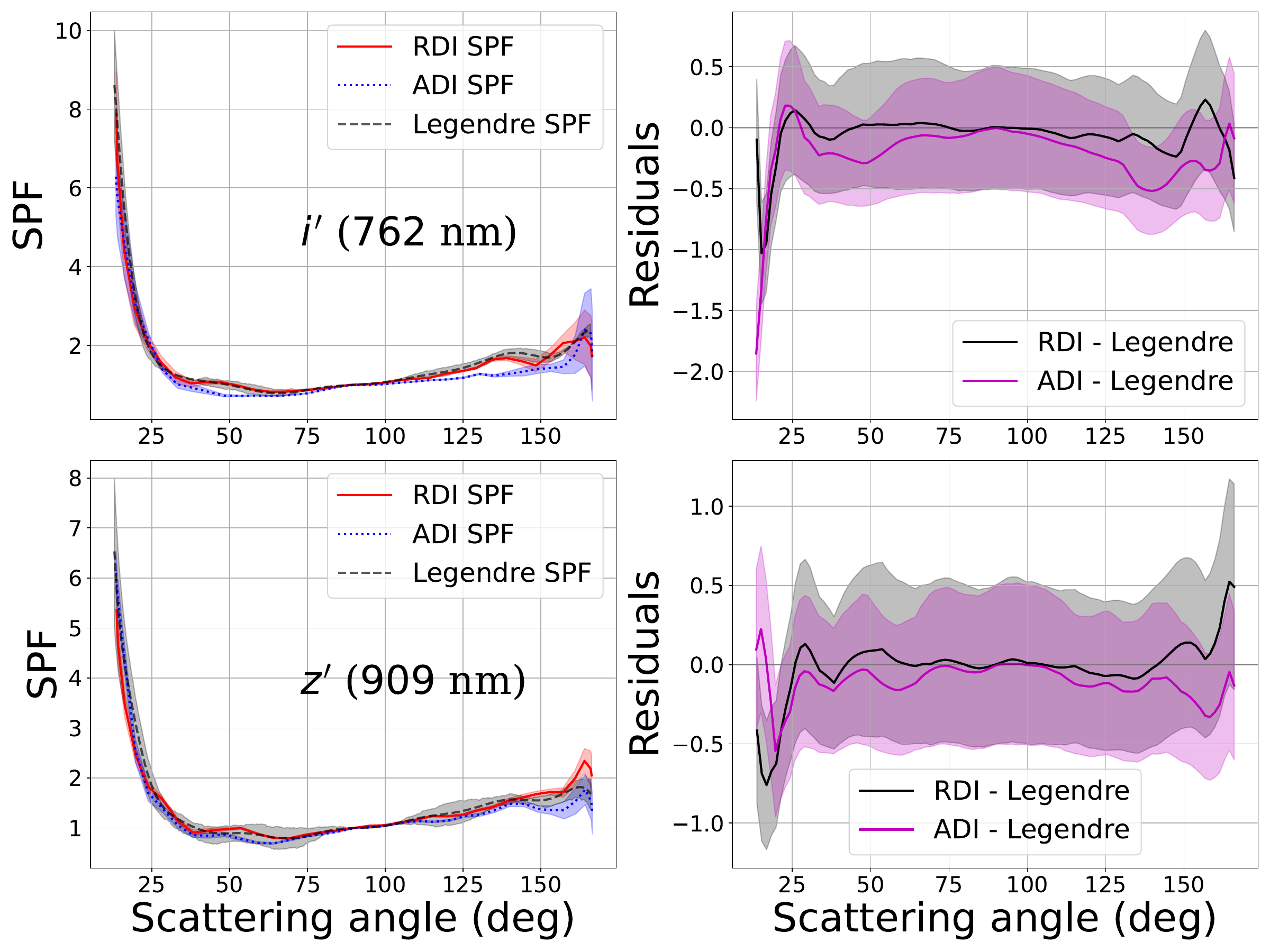}
    \caption{Results of inferring the azimuthal brightness distribution of the disk (i.e., the SPF) with our freeform models. Shaded regions denote 1 sigma uncertainties estimated from the spatial noise map mentioned in Section \ref{sec:setup}. \textbf{\textit{Top}:} Comparison of the \zband{} SPF measured using the freeform models for the cases of the \rdi{} image (red line) and the \adi{} image (dotted blue line). We include the SPF extracted using a parametric disk model that uses a basis of the Legendre polynomials to fit a custom SPF to the KLIP-reduced image as the gray line (see Kueny et al. 2025; submitted). The residuals between the freeform ADI and the custom Legendre-based SPF are shown in the right plot with the magenta line and the black line for the residuals between the freeform RDI and custom Legendre SPF. \textbf{\textit{Bottom:}} the same as the top plot but for the \zband{} images. The inconsistencies at the extrema of the scattering angle range are due to starlight leakage and the WDH contaminating different parts of the disk nearest to the IWA, as the \adi{} and \rdi{} datasets were taken across calendar years.}
    \label{fig:spf_comparison}
\end{figure*}

Figure \ref{fig:spf_comparison} shows the results of these efforts with 1-sigma uncertainties estimated from the spatial noise map described in Section \ref{sec:setup}. We note that the disagreement in the two SPFs (\iband{}ADI and SCL/Legendre) is only in the forward scattering peak, i.e., the smallest angles. This is likely because the KLIP reduction has under-subtracted starlight in this region, but the Legendre model is fitting that residual. The persisting stellar halo is easily seen in Figure \ref{fig:klip_comparison} in the top row where it is also unfortunately captured by our optimized freeform model (top middle row). This notion is supported by the similarity of the SPFs at the small scattering angles between the freeform RDI and custom Legendre-based flavors, as they were used to extract the SPF from the same \rdi{} image. We observe good agreement between the 3 SPF extractions for the \zband{} case (bottom two plots in Figure \ref{fig:spf_comparison}).

To gauge \texttt{ffortissimo}'s ability to recover a known SPF through artifacts induced by \adi{} and different noise profiles we performed an inject-and-recover assessment using a synthetic disk and two of our ``back-rotated" datasets (see Section \ref{fig:regularization}). To gain a better understanding of the influence of the WDH on the optimized freeform models, we used datasets both moderately-affected by the WDH (\iband{} \adi{} images from UT 2023 Mar. 09) and severely affected by the WDH (\rband{} \adi{} images from 2023 Mar. 12) in these injection and recovery tests. We show the results of these tests in Figure \ref{fig:inject_recover}. For the \iband{} dataset results (top row) we note good agreement between the SPFs extracted from the injected and recovered disk models with the exception of the highest scattering angles where signal-to-noise is lowest. For the \rband{} results (bottom row), we observe that some flux at the brightest part of the minor axis is missing in the recovered freeform model likely due to the strong WDH artifact plaguing this dataset (see Figure \ref{fig:residuals_comparison}). The results of these tests suggest that the fidelity at which we can reconstruct close-in disk features depends strongly on the presence of extended sources of noise such as the WDH. More work is needed to fully characterize the effect of systematic errors on our pipeline which we leave for a later paper.

\subsection{The potential for super resolution imaging}
\label{sec:deconvolution-subsec}

Because \texttt{ffortissimo} optimizes a model of a disk object (i.e., pre-convolution), one can infer spatial features in a disk image that are below the diffraction limit of the telescope.

We construct our model on approximately the same pixel grid as our detectors sample, which makes it over-sampled relative to the Nyquist-Shannon critical sampling frequency for all the filters used in MagAO-X. We neglect the small differences in pixel scale in the X- and Y-directions, discussed in detail in \cite{long_astrometric_2025}.

The risk, of course, is that a model fit with such features is only one of many possible models that---post-convolution---produce identical diffraction-limited forward-modeled images. This does not mean there is nothing to be learned from these models, provided we understand the relative (non-)uniqueness of a particular realization. We assume \textit{a priori} that speckle-dominated regions of the image closest to the host star will be poorly constrained. 

\begin{figure*}
    \centering
    \includegraphics[width=0.85\linewidth]{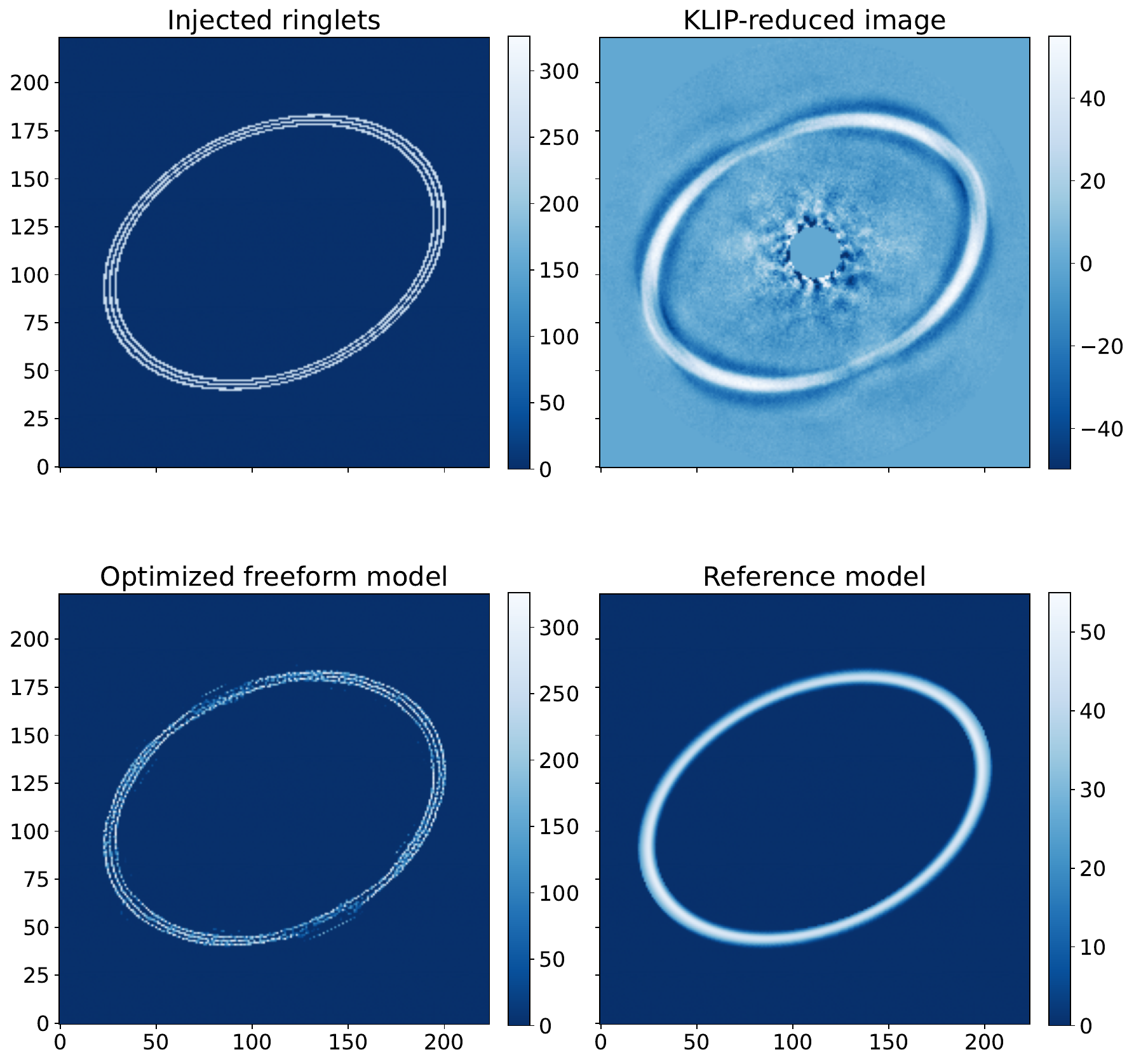}
    \caption{A demonstration of retrieval of object features by \texttt{ffortissimo} that are smaller than the diffraction limit of the telescope. We injected thin nested ringlets in a ``counter-rotated" \zband{} dataset ($\lambda/D = 2.4$ pixels) such that the \hr{} disk is averaged out. \textit{Top left:} injected nested ringlets with width and spacing of $\sim1$ pixel (0\farcsec012). \textit{Top right:} \adi{} reduced image of the injected ringlets. \textit{Bottom right:} The automatically-generated reference disk model used by \texttt{ffortissimo} to regularize the freeform model optimization. \textit{Bottom left:} recovered nested ringlets object. The results of this inject-and-recovery exercise suggest that our pipeline can infer features at spatial scales smaller than a resolution element.}
    \label{fig:hires}
\end{figure*}

This notion does come with caveats, however, especially given the current limitations of our modeling pipeline. Namely, the portion of the disk to be searched for small-scale features should be detected at high signal-to-noise and ideally far enough away from the speckle-dominated region closest to the host star. This reduces the likelihood of a feature being either learned speckle noise or a residual WDH artifact. To test these ideas, we injected an extended object with small spatial features into the ``back-rotated" set of the \zband{} \adi{} dataset (Figure \ref{fig:hires}). We simulated a face-on object with 3 nested thin, bright ringlets with width and spacing of $\sim1$ pixel (0\farcsec012). We convolved this object with the \zband{} PSF ($\text{FWHM}\approx 2.4$ pixels) and injected it into each of the coronagraphic science images prior to reducing the dataset with \adi{}. We observed a strong detection of the object, but the individual nested ringlets are clearly unresolved. We passed this reduced image and dataset into \texttt{ffortissimo} starting with a fitting region of uniform noise ($\mathcal{N}\sim[0,1]$) and an automatically-generated reference disk model (see Section \ref{sec:setup}) that bears no resemblance to the injected object other than the approximate width, position angle, and inclination. As shown in the bottom left panel in Figure \ref{fig:hires}, the final optimized freeform model recovers the nested ringlets across a substantial portion of the injected object, demonstrating \texttt{ffortissimo}'s ability to perform deconvolution on objects within the region being fitted.  As we improve our modeling pipeline's ability to recognize noise during model fitting (see Section~\ref{sec:limitations} below), inferring features on the smallest angular scales should become more reliable---especially in areas close to the star.

\begin{figure*}
    \centering
    \includegraphics[width=\linewidth]{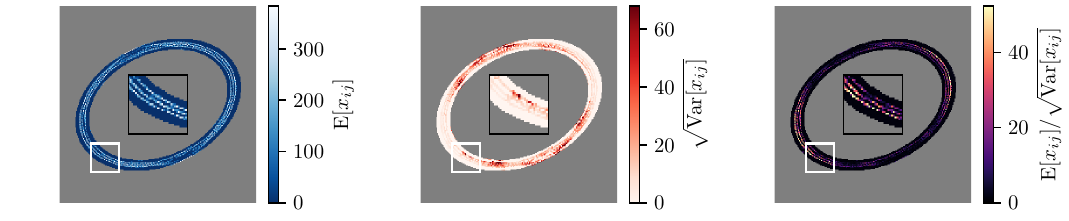}
    \caption{
    Left: the average converged model fit across all $N = 100$ runs with different random initial pixel values. Middle: the standard deviation of the final pixel values across all $N$ runs. Right: a ``signal-to-noise ratio'' quantifying how well-constrained a particular pixel's value is. When this value is low, the final value of that grid cell in the model is more dependent on its random initial value. The ``noise'' in this case is limited to summarizing the behavior of the model under different initial conditions, and should not be treated as a true signal-to-noise ratio.
    }
    \label{fig:uncertainties}
\end{figure*}

To estimate the pixel-wise uncertainty associated with the final optimized freeform model, we performed the aforementioned procedure across many separate optimization runs ($N = 100$) with randomized initial conditions and analyzed the statistics on this ensemble of final models. 
By varying the random seed used to initialize this first guess model, while holding all other parameters static, we can assess the final fit's sensitivity to the random initial conditions. 
The results of this experiment are showcased in Figure \ref{fig:uncertainties} where the left panel illustrates the average pixel value across trials, the middle panel the standard deviation of the pixel values, and the rightmost panel the average pixel values normalized by the standard deviation. One way to interpret the rightmost panel is that it is a measure of how consistently \texttt{ffortissimo} recovers the ringlets across realizations. We note that, in the regions closest to the major axis, the curvature of the ringlets is more or less in line with the direction of rotation, perhaps making these areas more robust against errors introduced by the image interpolation during rotation.


\section{Limitations and remaining challenges}
\label{sec:limitations}

\subsection{Noise characterization}
\label{sec:noise}

Currently, our freeform models are unable to efficiently distinguish faint disk features from the diffuse background of residual starlight (see Figures \ref{fig:klip_comparison} and \ref{fig:residuals_comparison}). One way this could be addressed is to enforce the expected post-KLIP background statistics on the residuals image when calculating the loss during the optimization loop. Getting an accurate measurement of the background statistics could be approached by aggregating reference PSF images from an archival database, as was performed by \cite{xie_reference-star_2022}, provided that the size and observational parameters are consistent with the target star observations.

Another source of noise that plagues high-contrast images of disks is the wind-driven halo (WDH; \citealt{cantalloube_wind-driven_2020}) which shows up in images as butterfly-shaped artifacts within the AO control region. Traditionally, high-pass filtering is employed to significantly reduce the footprint of this artifact in reduced images, however this operation would heavily alter the morphology of all but the highest inclination disks (i.e, edge-on). Based on our modeling trials thus far, the WDH is a nuisance that is learned by our freeform models as it, like the disk, is easily reconstructed using low spatial frequencies.

To draw quantitative conclusions from the freeform models, we need to understand to what extent the result is dependent on the randomized initialization of the model. The results are deterministic\footnote{The exact bit values from computation on accelerators like GPUs depend on things like driver versions and library versions. In practice, they cannot be relied on to remain \emph{exactly} equal over time---even on the same hardware.} but depend on a user-supplied random number generator ``seed''. To evaluate the sensitivity to initial conditions, we ran $N = 100$ different randomized states (seeds) on the Flatiron Institute computing cluster (Figure \ref{fig:uncertainties}).

Each grid point ran for 20 hours on a node with an NVIDIA RTX~6000~Pro~(Blackwell) GPU, which corresponded to $50\,000$ iterations. This is admittedly a large amount of computation, but we caution that the code has not been optimized for this use case. Also, our analysis suggests $\approx 1/10$ as many iterations per grid point would provide similar information. 

Given this limitation, we recommend using the current version of \texttt{ffortissimo} for exploratory morphological characterization. Any fine-scale features or signal inferred through PSF subtraction artifacts should be treated with caution, as the nature of deconvolution is that infinitely many equally valid solutions exist. This space of valid solutions is still constrained by the final image, however. So, while there may be infinitely many solutions, they will all share certain features.

The results using the current version of the pipeline presented in Section \ref{sec:performance} demonstrate \texttt{ffortissimo}'s potential for recovery of lost signals. However, we suggest that results from the current version are treated with caution and perhaps cross-validated with separate methods until the rigorous statistical framework needed for pixel-by-pixel variance is published.

\subsection{Complicated disk morphologies}
\label{sec:complicated_disks}

These validation trials and demonstrations have, so far, only included on-sky data pertaining to the iconic \hr{} disk. Despite a relatively complex brightness distribution owing to the resolved forward-scattering peak and asymmetry between the ansae, this disk presents as a clean, ring-like shape. Thus, a detailed performance analysis using \texttt{ffortissimo} on disks with complicated spatial features such as tightly-wound spiral arms (e.g., like the disk around SAO 206462) or nested ringlets (e.g., the disk around HD~141569A; \citealt{perrot_discovery_2016}) is subject to further study in a later paper. Concretely, \texttt{ffortissimo} can likely fit disk features of an arbitrary degree of complexity (see Figure \ref{fig:hires}), but regularization towards physical features in the models may suffer if the disk is not imaged at high $S/N$. We surmise that the main limitation will be the automated reference disk model fitting procedure mentioned in Section \ref{sec:setup} since fitting a simple model to systems with highly complex features is probably tricky. We will dedicate future work to exploring this issue and how it relates to the performance of the regularization. For instance, it could be the case that complicated disk features need only be partially modeled; the results from our tests where the forward-scattering peak in the reference \hr{}-like disk model being placed in a distinctly-different location (Figure \ref{fig:experiments}) and our simulations involving nested ringlets (Figure \ref{fig:hires}) are good evidence toward this conjecture.

\section{Summary and Future Work}
\label{sec:summary}

We developed a novel method to characterize circumstellar disks using KLIP-reduced images through a new Python pipeline \texttt{ffortissimo} that uses pixel-based freeform forward models. \texttt{ffortissimo} is open source and freely available on GitHub\footnote{\url{https://github.com/jkueny/debrisdisk_freeform_fit_and_plot}}. The methods utilized in the DiskFM module \citep{mazoyer_diskfm_2020} lie at the heart of \texttt{ffortissimo} to accurately recover disk features that have been distorted by artifacts induced by the PSF subtraction, including the forward-scattering peak. By constructing our freeform modeling pipeline within the framework of \textsc{Jax}, we make use of gradient descent optimization enabled by autodiff and GPU-accelerated array computations to optimize our freeform models, leading to dramatic time savings for model convergence. Using on-sky visible light images of the well-known \hr{} disk, we demonstrated that our freeform models can infer both a complex brightness distribution and realistic dust density profile. We also showed the potential for recovering spatial features in disks that are finer than the resolution element of the telescope.

While our freeform models facilitate characterization of highly complex disk features, separating noise from the true astrophysical signal is a remaining hurdle. Specifically, it is difficult for \texttt{ffortissimo} to separate noise from disk signal when updating the model parameters (i.e., pixels in the disk ROI) which leads to the model learning speckle noise and WDH artifacts especially at close IWAs. There would be value in more concentrated efforts to measure the noise contained within pixels that also contain disk flux to better inform the parameter weights. This could be done by, e.g., collecting data from a reference star which does not harbor a disk using the same observing strategy as the target star to enforce the expected background statistics in the residuals images. Additionally, \cite{males_mysterious_2021} highlighted the potential for using the WFS telemetry to estimate the speckle intensity at the focal plane which would be one avenue towards lessening the noise learned by our freeform models during optimization. Finally, the WFS telemetry could also be used to analyze preferential directions of residual wavefront error due to fast turbulence and the servo-lag inherent to AO instruments \citep{cantalloube_wind-driven_2020}, revealing the position angles of the WDHs in the raw images. This information could then be used to inform observing strategies that might mitigate this noise artifact in the final reduced images or perhaps be used in a parametric modeling routine that fits for WDHs in the raw images in post-processing for removal. We plan to include and build upon these ideas in experiments taking place in the near future.

\section{Acknowledgments}

We would like to thank the anonymous referee whose feedback vastly improved the quality of this manuscript.

J.K.K., J.R.M., and A.J.W. acknowledge support from the NSF, grant no. AST-2307613. J.D.L. was supported by the Flatiron Software Research Fellowship at the Flatiron Institute, a division of the Simons Foundation.

We are very grateful for support from the NSF MRI Award \#1625441. The Phase II upgrade program is made possible by the generous support of the Heising-Simons Foundation. The development of CACAO is supported by NSF Award \#2410616.

We thank Prof. Ewan Douglas and the University of Arizona Space Research Lab for access to computing resources during critical phases in our modeling. These computing resources were supported by funding by generous anonymous philanthropic donations to the Steward Observatory of the College of Science at the University of Arizona.

This work used Jetstream2 at Indiana University through allocation PHY250222 from the Advanced Cyberinfrastructure Coordination Ecosystem: Services \& Support (ACCESS) program, which is supported by National Science Foundation grant nos. 2138259, 2138286, 2138307, 2137603, and 2138296.

This material is based upon High Performance Computing (HPC) resources supported by the University of Arizona TRIF, UITS, and Research, Innovation, and Impact (RII) and maintained by the UArizona Research Technologies department.

\ This work has made use of data from the European Space Agency (ESA) mission
{\it Gaia} (\url{https://www.cosmos.esa.int/gaia}), processed by the {\it Gaia}
Data Processing and Analysis Consortium (DPAC,
\url{https://www.cosmos.esa.int/web/gaia/dpac/consortium}). Funding for the DPAC
has been provided by national institutions, in particular the institutions
participating in the {\it Gaia} Multilateral Agreement.

\clearpage

\bibliography{4F}{}
\bibliographystyle{aasjournal}

\end{document}